\documentclass[preprint,12pt]{elsarticle}
\usepackage[a4paper, margin=2.8cm]{geometry} 
\usepackage{amsmath}  
\usepackage{setspace}
\usepackage{empheq}
\usepackage{multirow}
\usepackage{booktabs}
\usepackage{tabularx}
\usepackage{siunitx}
\usepackage{makecell}
\usepackage{hyperref}
\allowdisplaybreaks
\usepackage{mathtools}
\usepackage[a4paper, margin=2.8cm]{geometry} %
\usepackage{graphicx}
\allowdisplaybreaks
\usepackage{mathtools}
\usepackage[center]{caption}
\usepackage{float} 
\usepackage{array}
\usepackage{subcaption}
\usepackage{float}
\usepackage{chngcntr}
\usepackage{xcolor}
\journal{....}

\begin{document}
\setstretch{1.3} 

\begin{frontmatter}

\title{Activation entropy helps explain anomalous flow stress temperature dependence in copper}

\author[inst1]{\corref{cor1}Mohammadhossein Nahavandian}
\author[inst1]{Liam Myhill}
\author[inst1,inst2]{\corref{cor1}Enrique Martinez}

\cortext[cor1]{Corresponding authors: enrique@clemson.edu; mnahava@clemson.edu}

\affiliation[inst1]{organization={School of Mechanical and Automotive Engineering},
            addressline={Clemson University}, 
            city={Clemson},
            postcode={29634}, 
            state={SC},
            country={United States}}
            
\affiliation[inst2]{organization={Department of Materials Science and Engineering},
            addressline={Clemson University}, 
            city={Clemson},
            postcode={29634}, 
            state={SC},
            country={United States}}    
            
\begin{abstract}
Thermal activation of dislocations is critical for predicting the mechanical response of materials under common experimental conditions. According to transition state theory (TST), the rate for the system to overcome free energy barriers depends on an attempt frequency, activation free energy, and temperature. We computed the rate for edge and screw dislocation dipoles to overcome their interaction fields at various temperatures, Langevin friction coefficients, and shear stresses using Molecular Dynamics (MD), Schoeck's entropy formalism and compared with Kramers rate theory \cite{kramers1940brownian}. Kramers theory matches the rates computed dynamically, which depend on Langevin friction, increasing with weaker friction. Statically, using Schoeck’s formalism to compute the entropy along the minimum energy path (MEP), we found significant entropic effects that lead to an increase of the critical resolved shear stress (\(\tau_{CRSS}\)) with temperature and could help explain the long-standing anomaly observed at low to intermediate temperatures in copper and other metals, where the flow stress increases with temperature.

\vspace{10pt}
\end{abstract} 

\begin{highlights}
\item Activation entropy helps explain anomalous flow stress temperature dependence in copper
\item Langevin friction affects the rate but does not change the energy barrier
\item Intermediate domain close to the \(\tau_{CRSS}\) at 0 K is dominated by entropy
\item Kramers' transition state theory reproduces atomistic rates dependence on Langevin friction
\end{highlights}


\end{frontmatter}

\vspace{10pt} 
\section{Introduction}
\label{sec:sample1}
Dislocations \cite{gilman1964influence,nahavandian2024design} are the primary carriers of permanent deformation in metals and alloys. To glide and accommodate deformation, dislocations need to overcome long and short-range interactions with other dislocations and obstacles of diverse nature \cite{hull2011introduction}. The role of dislocation lines on the permanent deformation process of metallic materials as a function of temperature and stress has been vastly studied. However, the influence of friction and thermal fluctuations on the rate at which dislocations overcome energy barriers is not fully understood. Moreover, many experimental studies have seen an abnormal flow stress increase with temperature in face-centered cubic (FCC) metals like Cu, Al, Ni, and Ag \cite{parkhomenko1982low}, which has not been fully explained. In the work by Turnage et al. \cite{turnage2018anomalous}, the normalized flow stress at 10\% strain in nanocrystalline Cu-\%10Ta shows an increase in flow stress with increasing temperature from 298 to 1073 K, highlighting the existence of a non-Arrhenius regime of deformation. In Cu-\%20Zn, an increase in critical stress for twining is reported at $50 <T < 300$~K  \cite{SCHNEIDER2021}. 
Takeuchi and Kuramoto \cite{takeuchi1973} show that the flow stress in Ni under different strain rates increases for $T<700 K$.  In Cu$_3$Au, the CRSS value increases for $T<660$~K \cite{Kuramoto01041976} due to cross-slip and diffusion processes that also lead to dynamic strain aging \cite{CAILLARD2001}. Parkhomenko and Pustovalov \cite{parkhomenko1982low} show that in Al-based alloys, there is an abrupt increase in yield stress at $T < 30$~K, and the same trend is observed in single-crystal Ag at $T < 75$~K. Also, Fusenig and Nembach \cite{FUSENIG1993} show that in Cu-Co alloys, an increasing trend in CRSS with $T < 200$~K occurs. 
Shim et al. \cite{SHIM2016276} state that this behavior at 200 K and below needs to be understood. To the best of our knowledge, none of the mentioned studies have found a clear answer to this behavior. In this work, we study thermally activated dislocation glide and observe that activation entropy can help explain this anomalous behavior of the flow stress.

In a seminal work, Granato and Luke \cite{granato1956application,granato1956theory} developed a theory of damping due to dislocations modeled as long vibrating strings \( L \) and line tension \( \Gamma \), and the frequency response is classified based on small and large damping effects. Relying on a modified Granato-Lücke \cite{granato1956theory} string model, Churochkin et al. \cite{churochkin2005low} studied the contribution to the low-frequency internal friction (IF) and the thermal conductivity of vibrating edge dislocation dipoles and showed that the presence of a reasonable density of vibrating dipoles provides a good fit to the thermal conductivity in superconducting samples, although, the internal friction experiments cannot be described within the standard fluttering string mechanism. Grabowski and Zotov \cite{grabowski2021thermally} studied thermally-activated screw dislocation dipole mobility in Nb using accelerated MD at constant temperature and strain. 
Bitzek and Gumbsch \cite{bitzek2004atomistic} conducted atomistic simulations to investigate the effects of drag and inertia on an accelerating edge dislocation. They determined the Peierls stress and the effective mass of an edge dislocation across various temperatures and stresses within a simple slab geometry. Cerrai and Freidlin \cite{cerrai2015large} established a large deviation principle for the Langevin equation with strong damping, describing the asymptotic behavior of the system as the damping parameter tends to zero. By simulating a 1-D shock impact problem, Blaschke and Luscher \cite{blaschke2021dislocation} assessed the significance of accurately modeling dislocation drag and dislocation density evolution under high-stress conditions. The simulations, which utilize a temperature, pressure, and character-dependent dislocation drag coefficient derived from first principles, indicate that the temperature dependence of dislocation drag results in a slight increase in the elastic precursor amplitude in single-crystal Al. In the study of IF of dislocations in ultra-low carbon bake-hardenable steel \cite{jung2014impulse}, the dislocation segment length is found to be the main parameter affecting the IF peak amplitude. In addition to metallic materials, the movement of the edge dislocation dipoles in 2D structures can affect the mechanical properties of planar graphene at high temperatures, where it is found that annihilation of dipoles with arm length $l > \SI{16}{\angstrom}$ can easily occur at high temperatures employing MD simulations \cite{galiakhmetova2024dynamics}. \\

In an author's recent study, it is shown that the activation entropy depends on both temperature and stress, and governs the rates for dislocations overcoming a long-range interaction with prismatic sessile loops at high stresses near \(\tau_{CRSS}\) \cite{NAHAVANDIAN2024112954}.
Wang and Cai \cite{wang2023stress} resolved the discrepancy between TST and MD in the prediction of cross-slip of screw dislocations, pointing out that the source of large activation entropy is from anharmonic effects, although they use harmonic transition state theory to estimate the rate. Also, it has been reported that the critical Escaig stress on the glide plane for dipole annihilation rapidly decreases from the 0 K value of approximately 400 MPa in face-centered cubic (FCC) Cu and Ni, making dipole annihilation nearly athermal at room temperature \cite{rao2015screw}.
A recent study using classical and machine-learned potentials shows a huge difference between harmonic and anharmonic approximation for the dislocation velocity \cite{allera2024activation}. Furthermore, entropic stabilization under tension and non-Arrhenius behavior at high temperatures was found for dislocation nucleation at free surfaces in Cu \cite{bagchi2025anomalous}. \\

One way to determine the role of Langevin friction in the dislocation glide mechanism is to compute the rate at which a dislocation overcomes an energetic barrier. In 1940, Kramers \cite{kramers1940brownian} employed the Langevin equation to compute the rate of the system over a barrier, analyzing a one-dimensional model of a particle jumping over a double-well potential. This analysis provided insight into how the surrounding environment affects the rate of reaction, where the reaction that takes place is that the particle overcomes local energy barriers to diffuse through the material. Kramers found that in the strong friction regime, friction restricts the movement of particles, slowing the reaction rate. In contrast, in the weak friction limit, an increase in friction enhances energy exchange with the medium, leading to a higher reaction rate.

The application of Kramers' theory to materials with defects remains relatively unexplored in the literature. Ferrando et al. \cite{ferrando1993kramers} investigated the Kramers problem by separating intra-well and inter-well dynamics in periodic potentials, focusing on jump rates and probability distributions. Using Fourier analysis and numerical methods with Klein-Kramers dynamics, they examined a range of damping conditions and potential barriers. Their study revealed significant deviations from expected behaviors and underscored the importance of finite-barrier corrections in underdamped regimes. Helium scattering has been used to measure Na atom diffusion on surfaces, with the dynamic structure factor as the key observable. By analyzing hopping distributions, Guantes et al. \cite{guantes2003kramers} applied Kramers’ turnover theory to deduce the physical properties of diffusing particles. The theory, incorporating finite-barrier corrections, was validated through Langevin equation simulations. Frank and Ricioaei \cite{frank2016reaction} introduced an extrapolation method, rooted in Kramers’ rate theory, to recover kinetics from potential-scaled MD simulations without the need to define reaction coordinates. Pollak and Ianconescu \cite{pollak2016kramers} studied Kramers turnover theory by introducing a finite barrier correction term, which accounts for the fact that the energy interval of the escaping particle is bounded from below. They tested this theory for motion on a cubic potential and relatively low barriers. The recent study by Denton et al. \cite{denton2024influence} investigated the effect of friction coefficient on Langevin dynamics by the ISOKANN algorithm, which combines the transfer operator approach with machine learning techniques. We have not been able to find a systematic investigation of the role of friction in the rates of dislocations overcoming energy barriers in the context of Kramers' theory. \\

In this paper, we perform atomistic simulations for FCC copper (Cu) including dislocation dipoles of both edge and screw characters at different temperatures and applied shear stresses. We analyze the rate for these dislocation dipoles to cross each other over the elastic interaction barrier using various Langevin frictions. We studied a range of different applied shear stresses and compared the results with theoretical approaches based on rate predictions through TST \cite{NAHAVANDIAN2024112954} and Kramers rate theory \cite{kramers1940brownian,hanggi1990reaction}. Furthermore, we also show that entropic effects lead to an increase in critical resolved shear stress, shedding light on the origin behind the anomalous behavior of the flow stress observed in previous experimental studies.

\section{Methods}
\label{sec:methods}
\subsection{Simulation cell configuration}
\begin{figure}[H]
    \centering
    \begin{subfigure}[b]{0.45\textwidth}
        \centering
        \includegraphics[width=\textwidth]{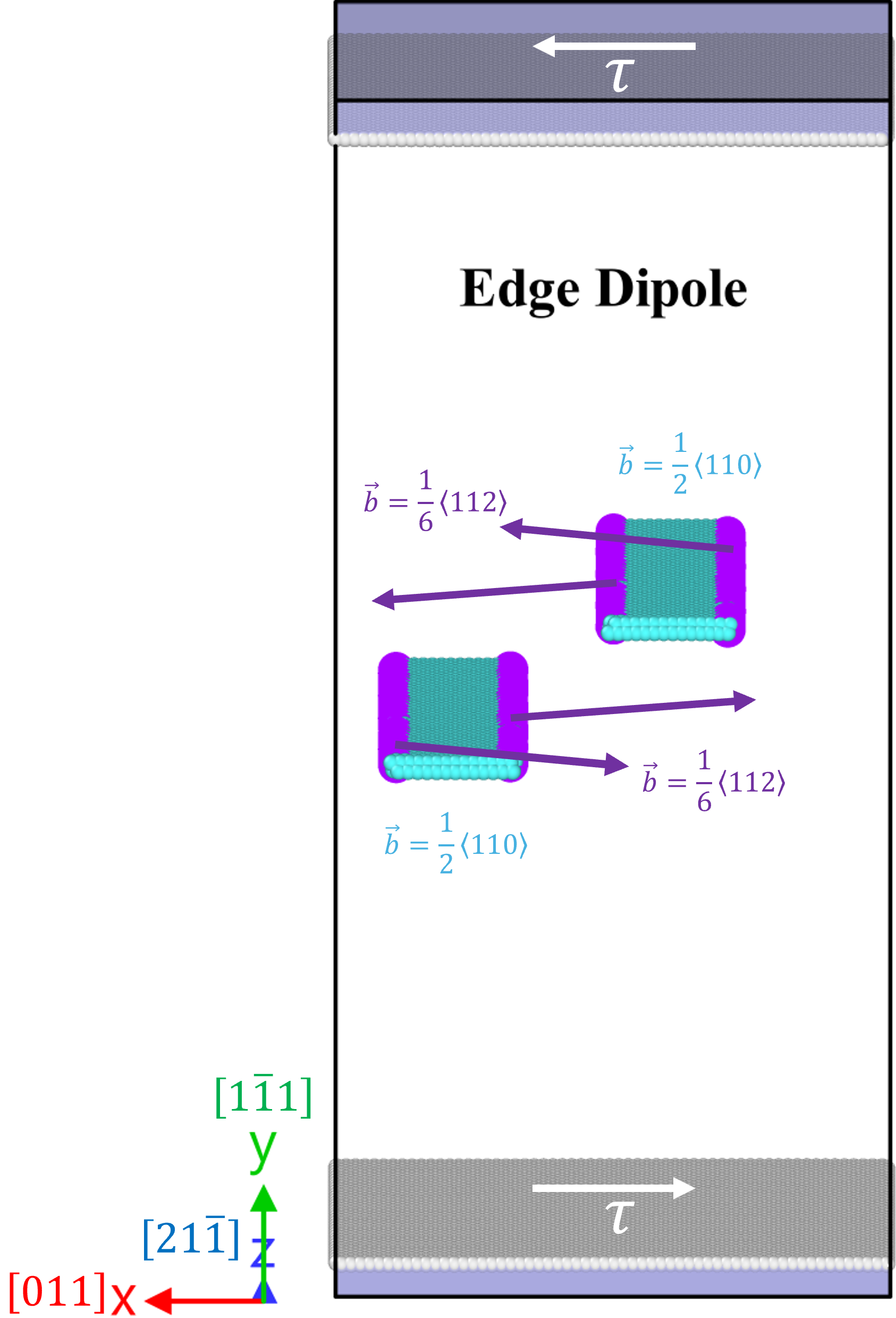}
        \caption{}
        \label{fig:edgeconf}
    \end{subfigure}
    \hfill
    \begin{subfigure}[b]{0.45\textwidth}
        \centering
        \includegraphics[width=\textwidth]{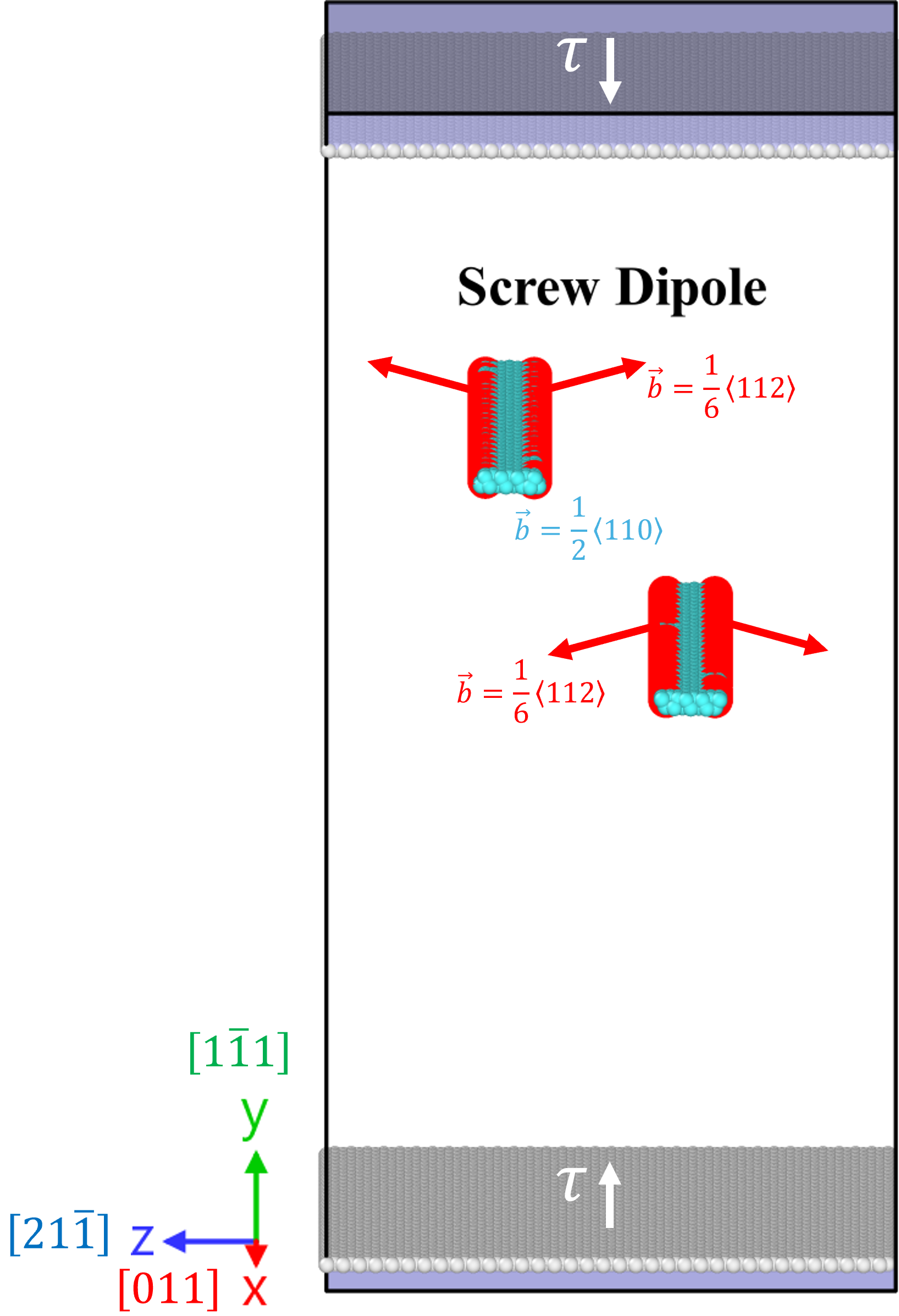}
        \caption{}
        \label{fig:screwconf}
    \end{subfigure}
    \caption{Simulation cells including dislocation dipoles, each dissociated to two Shockley partials (a) edge dislocation (b) screw dislocation. Only atoms in a non-FCC crystalline structure are shown.}
    \label{fig1}
\end{figure}

We perform atomistic simulations using LAMMPS \cite{plimpton1995fast} under various stresses, temperatures, and Langevin damping coefficients. The simulation samples contain over 600,000 Cu atoms, including either an edge or a screw dislocation dipole with orientation $x=[011]$, $y=[1\bar{1}1]$, and $z=[21\bar{1}]$. The edge dislocations align with the $z$ direction, while screws are parallel to the $x$ direction with lengths \SI{154.820}{\angstrom} and \SI{153.177}{\angstrom}, respectively (see Fig.\ref{fig1}). We used the Atomsk code \cite{HIREL2015212} to include the dislocations relying on anisotropic elasticity for the atomic displacements. The dislocation dipoles are separated in the $y$-direction by \SI{37.541} and \SI{60.878}{\angstrom} (distance between glide planes). The sample dimensions, provided in Table \ref{table1}, are periodic in $x$ and $z$ directions and free in $y$ induced by a buffer layer of \SI{20}{\angstrom} in thickness.
 
\begin{table}[H]
\centering
\caption{Simulation box dimensions (\AA)}
\small
\begin{tabular}{c c c c c c c}
\hline
Included defect &  $x_{\text{min}}$ & $x_{\text{max}}$ & $y_{\text{min}}$ & $y_{\text{max}}$ & $z_{\text{min}}$ & $z_{\text{max}}$ \\ \hline
Edge       &0.21          & 150.38         & -11.27         & 329.34         & 0.39           & 154.17        \\ \hline
Screw      &-0.03          & 154.90        & -10.07         & 326.95         & 0.06          & 153.24         \\ \hline
\label{table1}
\end{tabular}
\end{table}

\subsection{Molecular Dynamics}
We used Mishin et al. \cite{mishin2001structural} interatomic potential for Cu to run MD under several shear stresses and a temperature range from 300 to \SI{700}{\kelvin}. To control the temperature, we use a Langevin thermostat with different damping coefficients varying between 0.1 to \SI{10}{\pico\per\second} (denoted by \(\gamma\)) and corresponding to characteristic relaxation times of 10 to \SI{0.1}{\pico\second}, respectively, to study its effect on transition rates. We visualize the thermally activated processes in Ovito \cite{stukowski2009visualization,stukowski2010extracting,stukowski2012elastic}, and use a common neighbor analysis (CNA) and the dislocation extraction algorithm (DXA) to detect and characterize dislocation lines within the atomistic crystal structure. We computed the rate for dislocations to cross the energetic barrier created by dislocation-dislocation interactions as the inverse of the average waiting times for multiple realizations at each stress, temperature, and damping coefficient for a total of 465 simulations.

\subsection{Schoeck's Entropy}
As we have discussed in a previous work \cite{NAHAVANDIAN2024112954}, in 1980 Schoeck \cite{schoeck1980entropy}, using the theory of elasticity for finite strains, developed an expression for the change in entropy induced by internal strains (Eq.\eqref{eq:Schoeck}) containing a term on hydrostatic strains and the thermal expansion coefficient, and another term related to the dependence of the elastic constants on temperature.

\begin{equation}
    \Delta S = \alpha_V K \int _{\Omega}  V_{ii} d\Omega - \frac{1}{2} \frac{\partial C_{iklm}}{\partial T} \int _{\Omega}  V_{ik} V_{lm} d\Omega ,
\label{eq:Schoeck}
\end{equation}
where \(\Delta S\) is the change in entropy, \(\alpha_V\) the volumetric thermal expansion coefficient, \(V_{ik}\) the elastic strain components of the second order Green's strain tensor, \(C_{iklm}\) the \(4^{th}\)-order elastic constant tensor, and \(\Omega\) the volume. To track the time evolution (dissipation) of entropy, given that the elastic constants, volumetric thermal expansion, and bulk modulus are time-independent, the time derivative of Schoeck's specific entropy over the whole domain can be expressed as follows:
\begin{equation}
    \Delta \dot{S} = \alpha_V K \int _{\Omega}  \dot{V}_{ii} d\Omega - \frac{1}{2} \frac{\partial C_{iklm}}{\partial T} \int _{\Omega}  ( \dot{V}_{ik} V_{lm} + V_{ik} \dot{V}_{lm}) d\Omega ,
\label{eq:Schoeck_time}
\end{equation}
which represents the entropy dissipation in a specific process where internal strains vary in time. Following the Schoeck approach, if we expand the free energy up to \(4^{th}\) order, Eq.(\ref{eq:Schoeck}) can be generalized as:
\begin{equation}
    \Delta S_{4^{th}} = \alpha_V K \int _{\Omega}  V_{ii} d\Omega - \frac{1}{2}  \int _{\Omega} \frac{\partial C_{iklm}}{\partial T}  V_{ik} V_{lm} d\Omega  - \frac{1}{6} \int _{\Omega} \frac{\partial ^2 C_{iklm}}{\partial V_{pq}\partial T} V_{ik} V_{lm} V_{pq} d\Omega ,
\label{eq:Schoeck_4th}
\end{equation}
where Eq.(\ref{eq:Schoeck_4th}) can be useful when the elastic constant is both function of temperature and strain.

We computed the thermal expansion coefficient and the elastic constants depending on temperature with the given interatomic potential \cite{mishin2001structural} (see Table \ref{elastic properties} and \ref{App: elastic_constant}):
\begin{table}[H]
\centering
\caption{Elastic Properties for Cu}
\resizebox{0.55\textwidth}{!}{
\begin{tabular}{c|c}
\hline
\textbf{Parameter} & \textbf{Expression (\SI{}{\giga\pascal})} \\
\hline
\(C_{11}\) & $5.263 \times 10^{-6} T^2 - 0.014T + 168.200$ \\
\hline
\(C_{12}\) & $1.009 \times 10^{-5} T^2 - 0.004 T + 121.000$ \\
\hline
\(C_{44}\) & $-4.885 \times 10^{-6} T^2 - 0.003T + 73.73$ \\
\hline
\(\alpha_{V}\) & $24 \times 10^{-9} T + 3 \times 10^{-5}$ \\
\hline
\(K\) & $-2.502 \times 10^{-4}T + 135.53$ \\
\hline
\end{tabular}
}
\label{elastic properties}
\end{table}

We computed the MEP using the nudged elastic band (NEB) \cite{henkelman2000climbing} method for different applied shear stresses, with component $\tau _{xy}$ for the edge and $\tau _{yz}$ for the screw. One key step in this process is the consideration of the thermal expansion prior to the calculation of the MEP. As such, we observe an elongation in the $y$ direction that we took into account to obtain the MEP. Since the position of the saddle point is unknown, the number of replicas along the MEP needs to be large enough not to skip the maximum. In principle, the reaction coordinate for the saddle point in the free energy landscape might not correspond with the maximum of enthalpy \cite{bagchi2025anomalous}. This calculation will result in the enthalpy for each replica along the path. Once we have the atomic configurations along the MEP we can follow the same methodology as in Ref. \cite{NAHAVANDIAN2024112954} to compute the activation entropy. Taking the first replica, i.e. the initial state, as the reference, we compute the atomic strains. The atomic volume is computed for every atom using a Voronoi tessellation \cite{stukowski2009visualization} excluding two atomic planes at the surfaces normal to the \(y\) direction and the integrals in Eq. \ref{eq:Schoeck} are discretized to obtain the change in entropy.
With the enthalpy and entropy along the MEP, we can find the Gibbs free energy \(\Delta G = \Delta H - T \Delta S\). The activation enthalpy contains a constant term for zero applied stress and the elastic and plastic work done by the applied stress.

According to TST, the transition rate is given by
\begin{equation}
    \begin{split}
    \Gamma_{TST} & = \nu \exp\left(-\frac{\Delta G_a }{k_BT}\right) = \nu \exp\left(\frac{\Delta S_a^h}{k_B}\right)\exp\left(-\frac{ \Delta H_a -T\Delta S_a^{nh}}{k_BT}\right) \\
    & =\nu \exp\left(\frac{\Delta S_a^h}{k_B}\right)\exp\left(-\frac{ \Delta G_a^{nh}}{k_BT}\right),
    \end{split}
\label{eq:TST}
\end{equation}
where \(k_B\) is the Boltzmann constant and  \(T\) the temperature. \(\nu\) is an attempt frequency, and \(\Delta G_a\), $\Delta H_a$, $\Delta S_a^h$ and $\Delta S_a^{nh}$ are the activation free energy, activation enthalpy, activation harmonic entropy and activation anharmonic entropy, respectively, and $\Delta G_a^{nh}=\Delta H_a -T\Delta S_a^{nh}$ is the anharmonic component of the activation free energy.

\subsection{Kramers Rate Theory}
H. A. Kramers' (1940) \cite{kramers1940brownian,hanggi1990reaction} formulated transition rates for one-dimensional systems starting from the Langevin equation. According to Kramers' \cite{hanggi1990reaction}, the rate of a process overcoming a barrier is given by:
\begin{equation}
    \Gamma_{Kr} = \frac{\gamma}{\omega_B} \left( \sqrt{\frac{1}{4} + \frac{\omega_B^2}{\gamma^2}} - \frac{1}{2} \right) \cdot \frac{\omega_A}{2\pi} \exp\left(-\frac{ \Delta H_a -T\Delta S_a^{nh}}{k_BT} \right),
\label{eq:kramers}
\end{equation}
where $\gamma$ is the Langevin friction, and $\omega_A$ and $\omega_B$ are angular frequencies described below, and where the prefactor is assumed to be harmonic with:
\begin{equation}
    \nu \exp\left(\frac{\Delta S_a^h}{k_B}\right) = \frac{\gamma}{\omega_B} \left( \sqrt{\frac{1}{4} + \frac{\omega_B^2}{\gamma^2}} - \frac{1}{2} \right) \cdot \frac{\omega_A}{2\pi}.
\label{eq:prefactor_kramers}
\end{equation}

In the low friction regime \(\gamma \ll \omega_B\), \(\Gamma_{Kr}\) equals \(\Gamma_{TST}\):

\begin{equation}
    \Gamma_{Kr} = \Gamma_{TST} = \frac{\omega_A}{2\pi} \exp \left (- \frac{ \Delta H_a -T\Delta S_a^{nh}}{k_BT}\right ),
\label{eq3}
\end{equation}
while under high friction \(\omega_B\ \ll \gamma\):

\begin{equation}
    \Gamma_{Kr} = \frac{\omega_A \omega_B}{2\pi \gamma} \exp\left (-\frac{ \Delta H_a -T\Delta S_a^{nh}}{k_BT}\right).
\label{eq4}
\end{equation}

For the process at study in this work, the rate prefactor $\nu(L)$ is proportional to the dislocation length $L$ and is expressed as \(\nu(L) = \nu_\text{Kr} \frac{L}{b}\), where $\nu_\text{Kr}$ represents an attempt frequency given by:
\begin{equation}
    \nu_{Kr} = \frac{\gamma}{\omega_B} \left( \sqrt{\frac{1}{4} + \frac{\omega_B^2}{\gamma^2}} - \frac{1}{2} \right) \cdot \frac{\omega_A}{2\pi} ,
\label{eq:kramers prefactor}
\end{equation}
and $b$ denotes the magnitude of the dislocation Burgers vector, presumed to be the smallest repeat distance along the dislocation line that can hop over the barrier. \(\omega_A\) and \(\omega_B\) are the angular frequencies depending on the curvature of the potential energy function at initial and saddle points, respectively. 
\begin{equation}
   \omega_A = \sqrt{\frac{1}{m} \left( \frac{d^2 \Delta H}{dx^2} \right)_{x=x_A}}   
\label{eq5}
\end{equation}
\begin{equation}
    \omega_B = \sqrt{\frac{1}{m} \left( \left| \frac{d^2 \Delta H}{dx^2} \right| \right)_{x=x_B}}
\label{eq6}
\end{equation}
with \(m\) the particle mass, which in this case we take as the total mass of atoms in the simulation cell.

Hence we can re-write Eq.\eqref{eq:kramers} as:
\begin{equation}
    \Gamma_{Kr} = \nu_{Kr} \frac{L}{b} \exp\left(-\beta \Delta G_a^{nh} \right),
\label{eq:kramerscorrect}
\end{equation}

\section{Results and Analysis}
\begin{figure}[H]
    \centering
    \includegraphics[width=12cm]{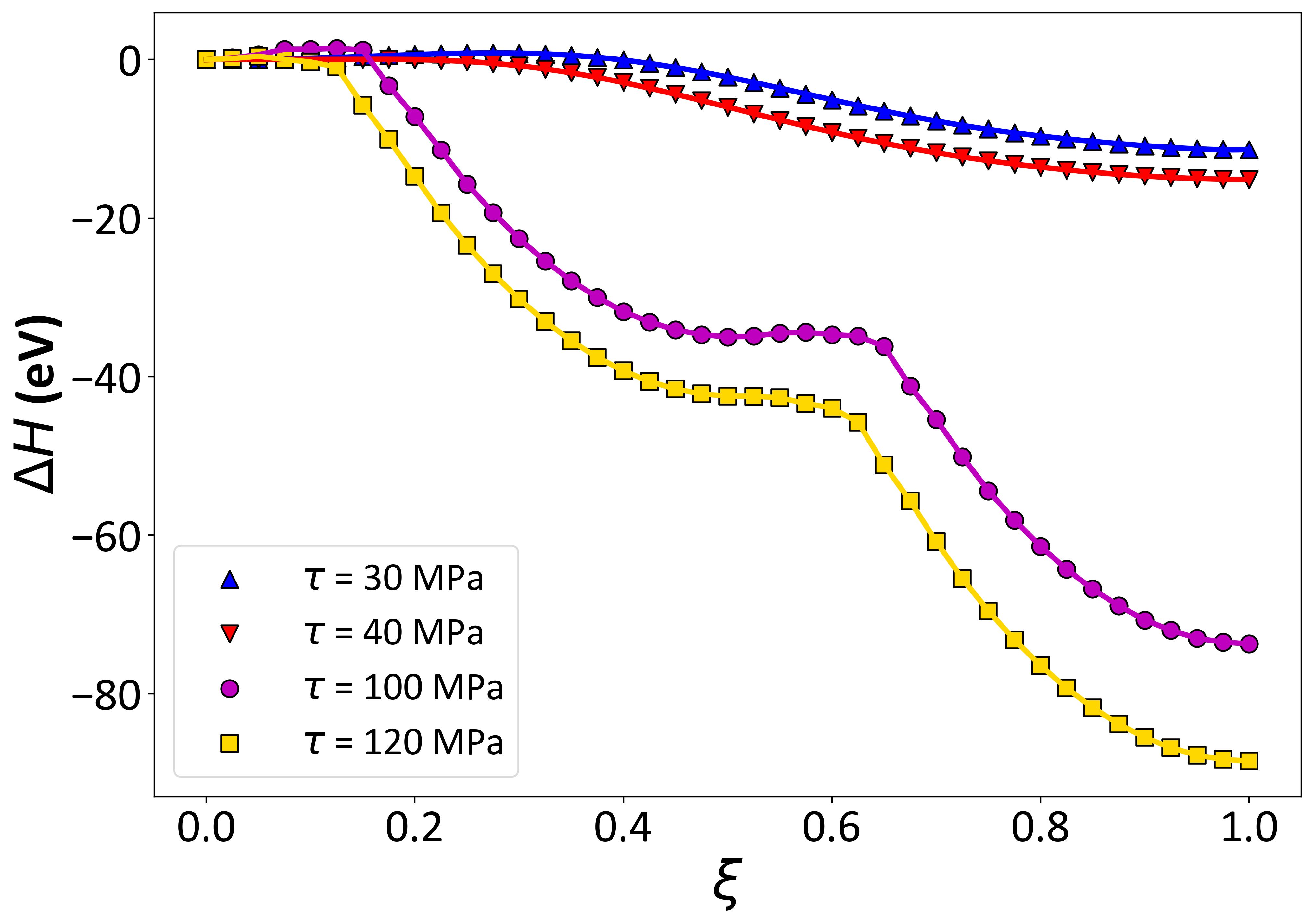} 
    \caption{Enthalpy along the NEB coordinate for cell including screw and edge dislocation dipole}
    \label{fig:enthalpy1}
\end{figure}

\begin{figure}[H]
    \centering
    \begin{minipage}{0.48\textwidth}
        \centering
        \includegraphics[width=\textwidth]{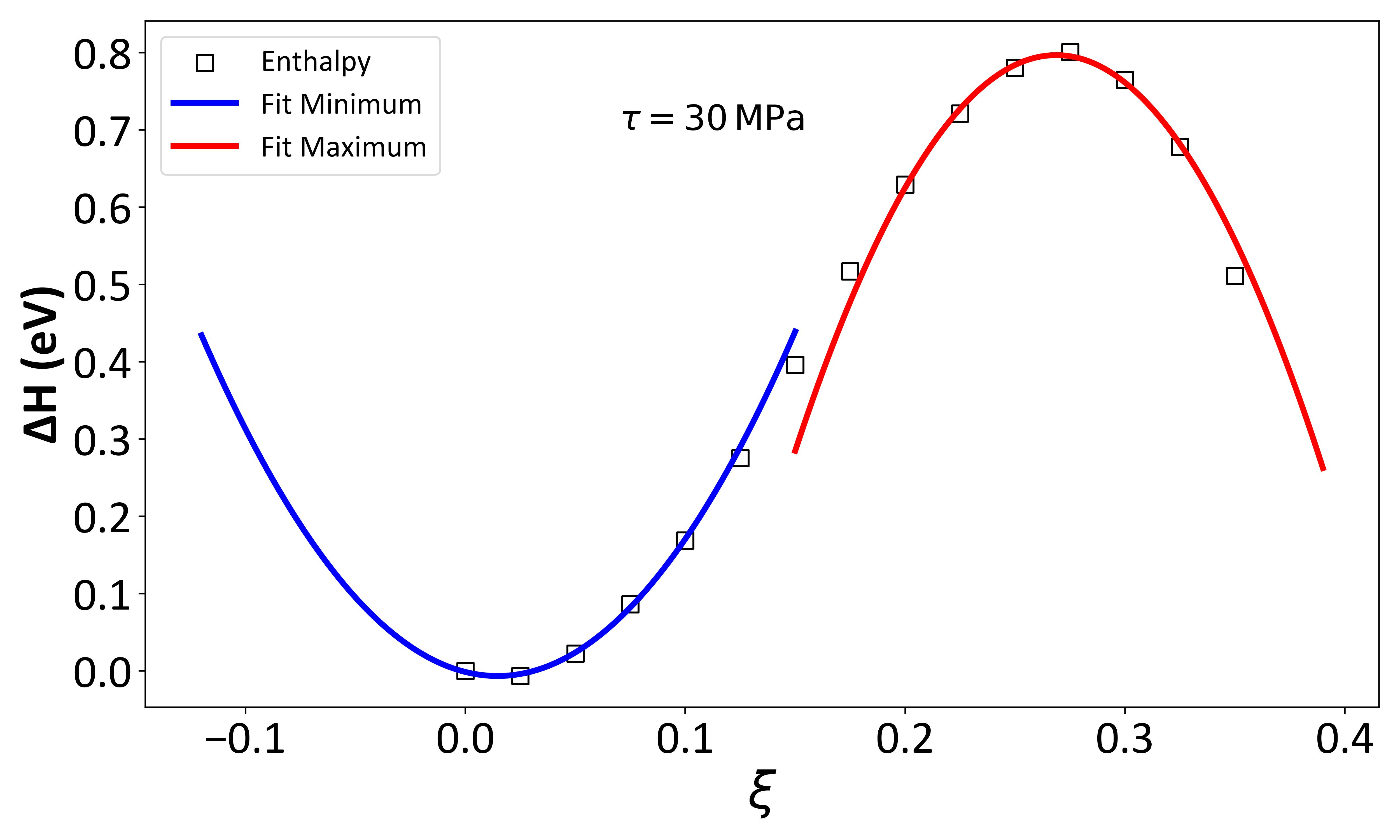}
        \subcaption{}
        \label{fig:BarrierKr30}
    \end{minipage}
    \hfill
    \begin{minipage}{0.48\textwidth}
        \centering
        \includegraphics[width=\textwidth]{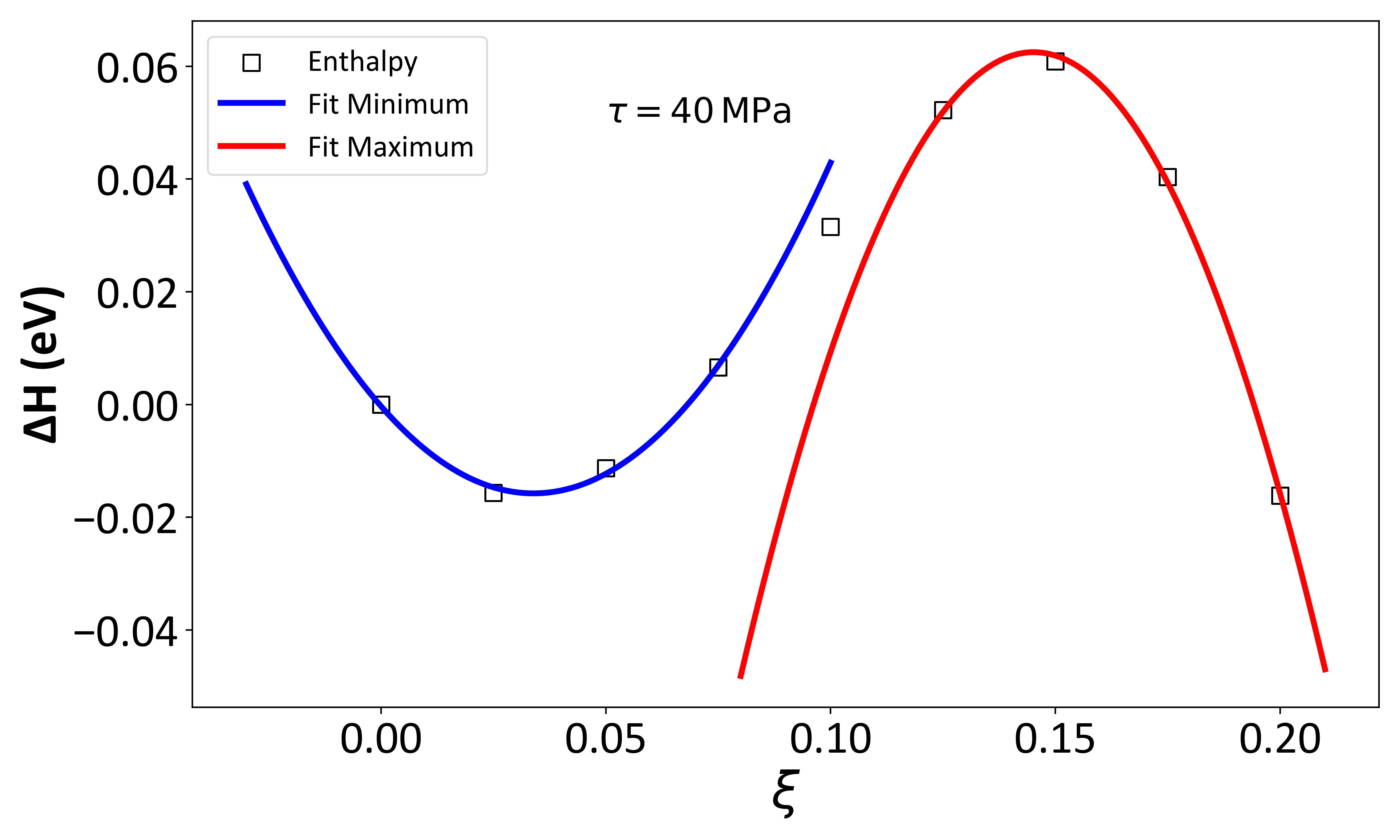}
        \subcaption{}
        \label{fig:BarrierKr40}
    \end{minipage}
    \vfill
    \begin{minipage}{0.48\textwidth}
        \centering
        \includegraphics[width=\textwidth]{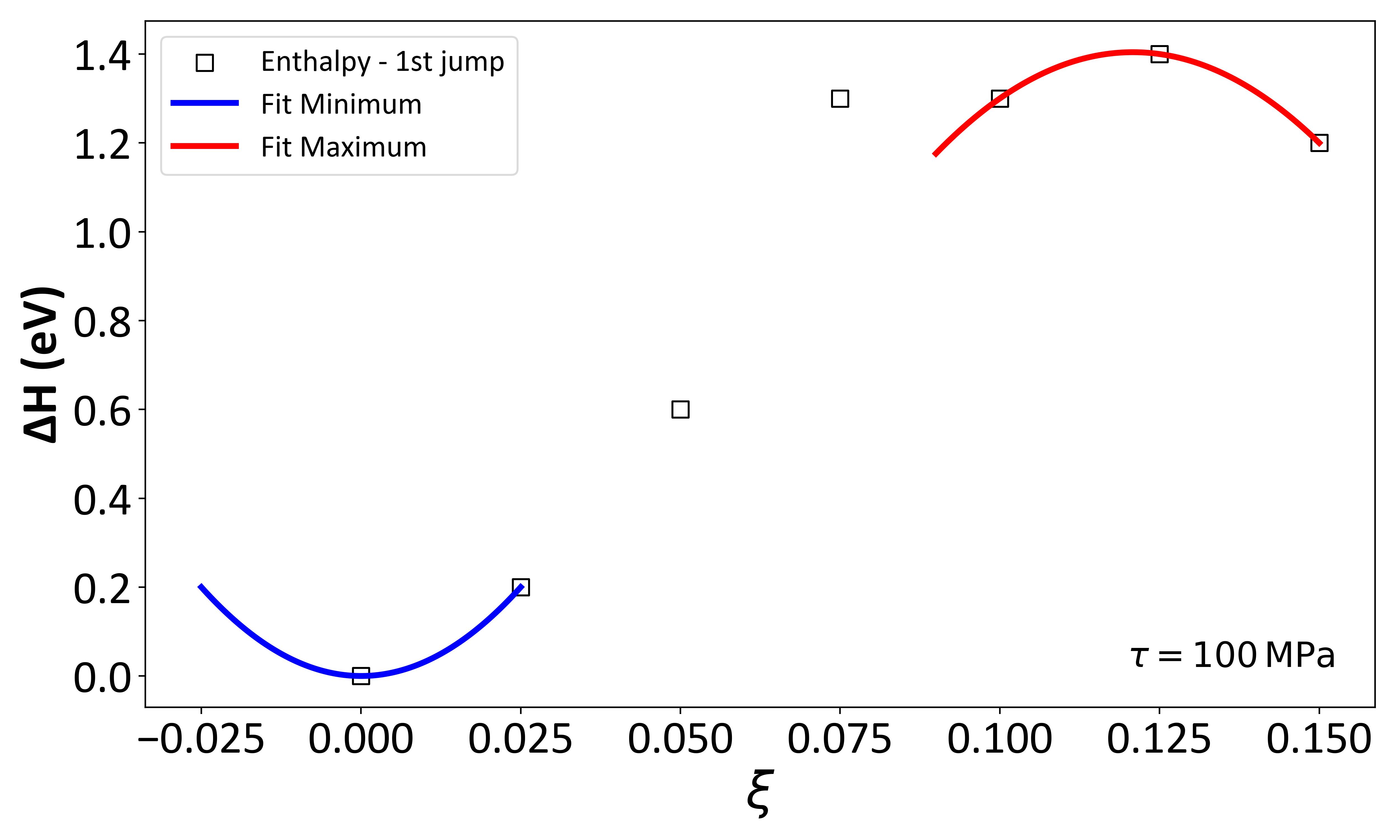}
        \subcaption{}
        \label{fig:BarrierKr1001}
    \end{minipage}
    \hfill
    \begin{minipage}{0.48\textwidth}
        \centering
        \includegraphics[width=\textwidth]{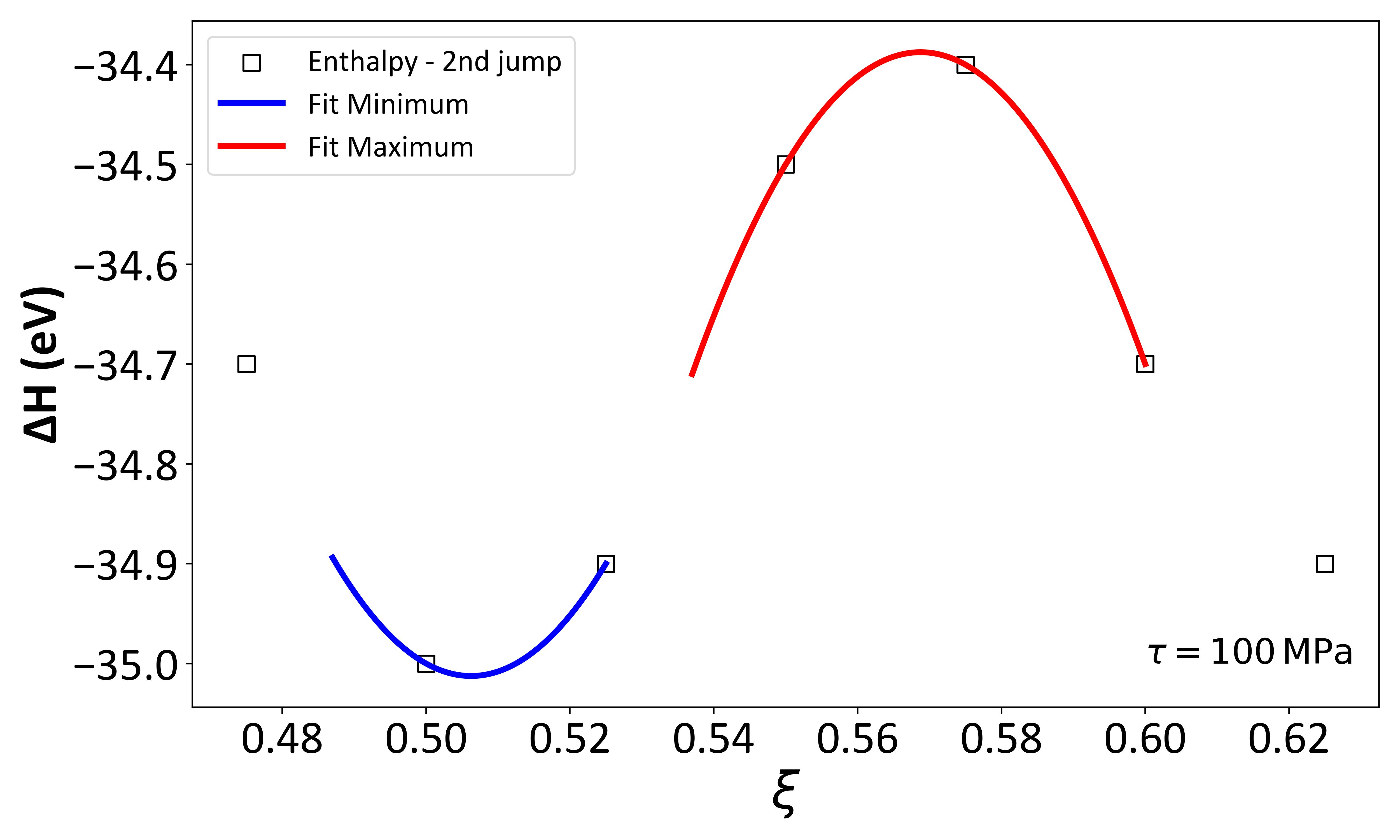}
        \subcaption{}
        \label{fig:BarrierKr1002}
    \end{minipage}
    \vfill
    \begin{minipage}{0.48\textwidth}
        \centering
        \includegraphics[width=\textwidth]{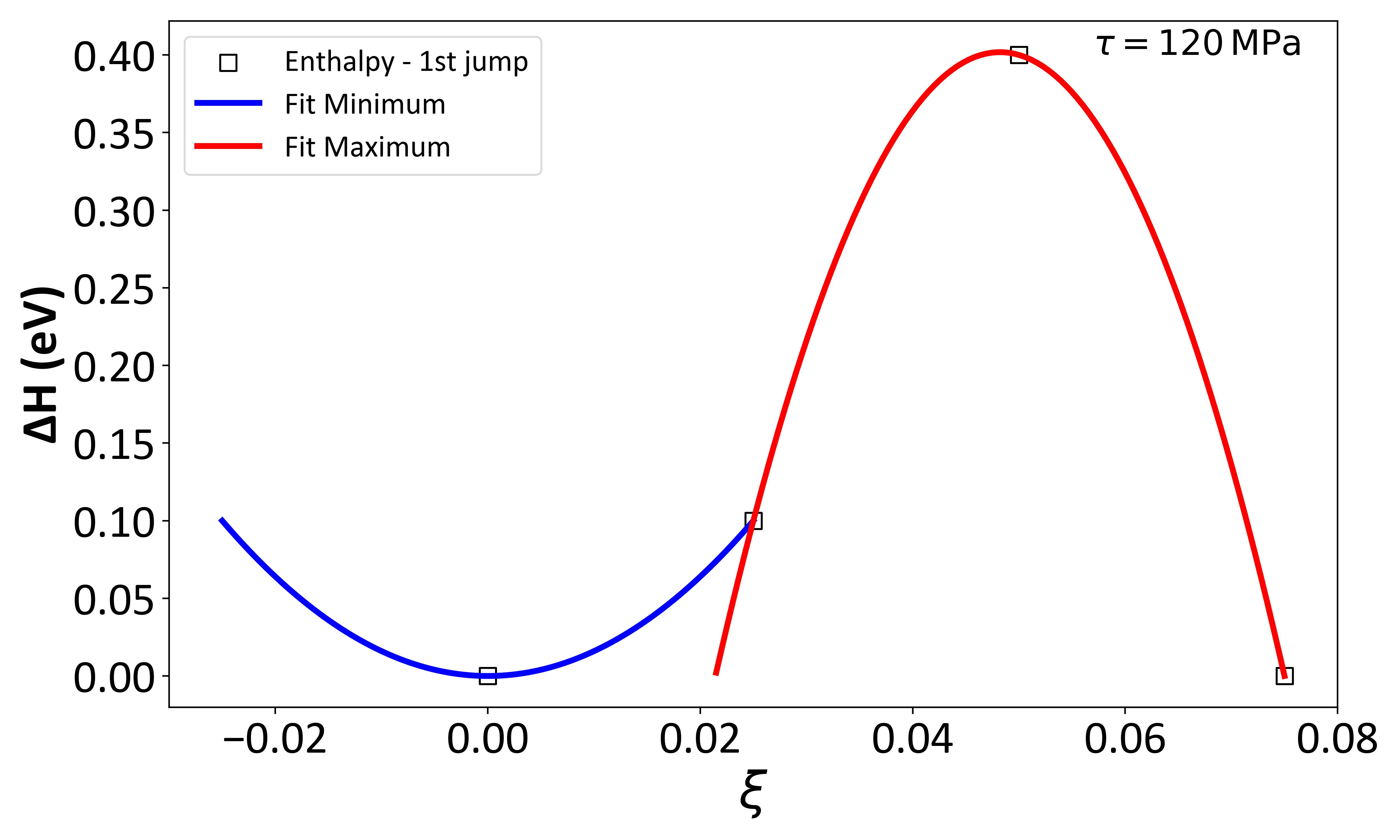}
        \subcaption{}
        \label{fig:BarrierKr1201}
    \end{minipage}
    \hfill
    \begin{minipage}{0.48\textwidth}
        \centering
        \includegraphics[width=\textwidth]{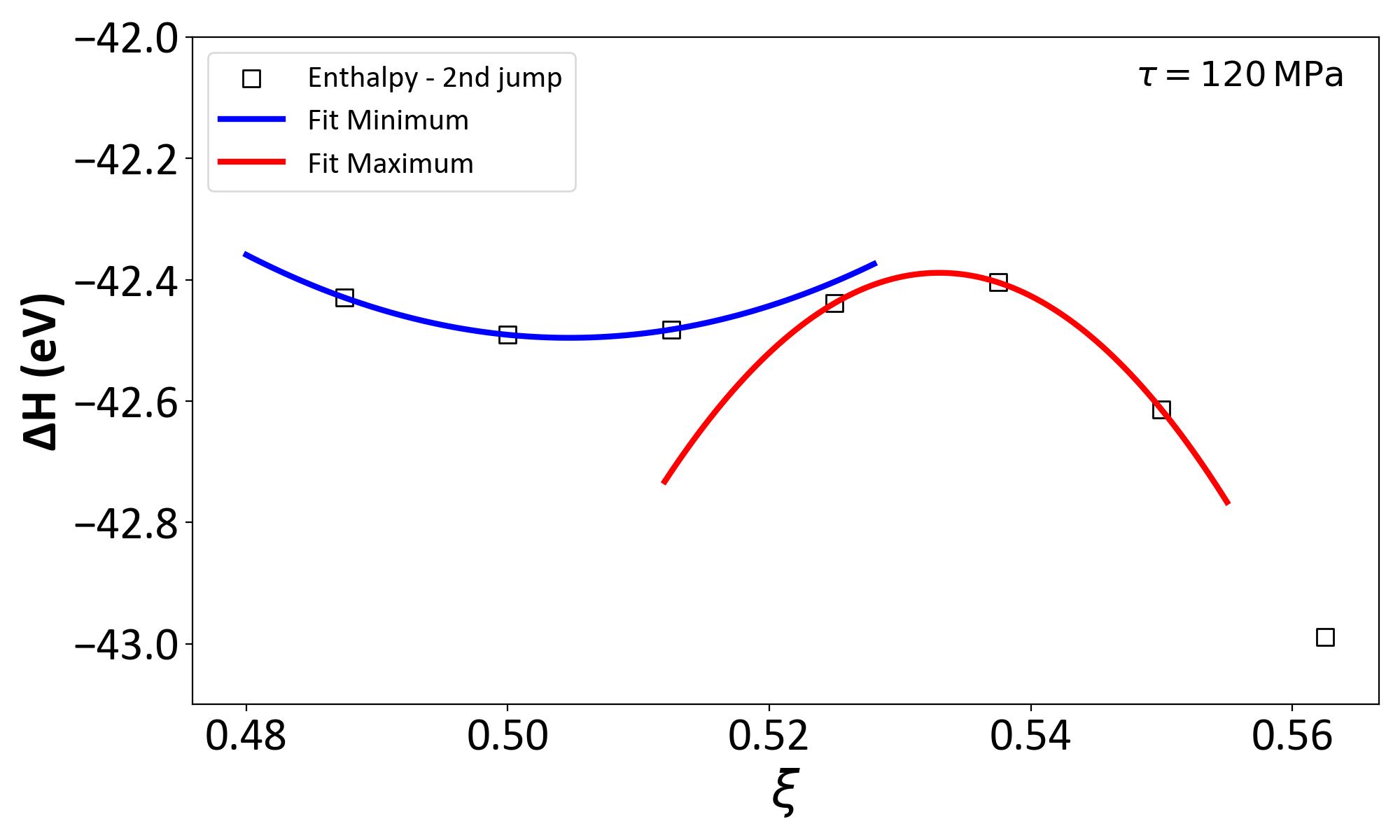}
        \subcaption{}
        \label{fig:BarrierKr1202}
    \end{minipage}
    \caption{Barrier along the MEP with fits at minima and saddle points for cell including screw dislocation dipole at a) \SI{30}{\mega\pascal} b) \SI{40}{\mega\pascal}, and edge dislocation dipole at c) first jump \SI{100}{\mega\pascal} d) second jump \SI{100}{\mega\pascal} e) first jump \SI{120}{\mega\pascal} f) second jump \SI{120}{\mega\pascal}. Blue solid line is harmonic fit at minimum and red solid line is harmonic fit at saddle point.}
    \label{fig:enthalpy}
\end{figure}

First, we computed the enthalpy along the MEP at different stress levels for both types of dislocations (shown in Fig. \ref{fig:enthalpy1}), where an increase in stress results in a reduction in the enthalpy barrier.  
We have fitted quadratic functions to the minimum and saddle point regions of the MEP. Table \ref{Potential Energy} shows the functions for the screw dipole at \SI{30}{\mega\pascal} and \SI{40}{\mega\pascal} and for the edge dipole at \SI{100}{\mega\pascal} and \SI{120}{\mega\pascal}, as a function of the reaction coordinate \(\xi\). These functions are plotted in Fig. \ref{fig:enthalpy}.

\begin{table}[H]
\centering
\caption{Enthalpy as a function of the reaction coordinate \(\xi\) at initial (i) and saddle point (\(\dagger\))}
\resizebox{1\textwidth}{!}{
\begin{tabular}{c|c|c}
\hline
\textbf{Stress value} & \textbf{\(\Delta H^{\dagger}(\xi)\) (\SI{}{eV})} & \textbf{\(\Delta H^{i}(\xi)\) (\SI{}{eV})} \\
\hline
\(\tau = \SI{30}{\mega\pascal}\) & \makecell{$-36.33 \xi^2 + 19.53 \xi - 1.827$} & \makecell{$24.33 \xi^2 - 0.7118 \xi - 0.001478$}  \\
\hline
\(\tau = \SI{40}{\mega\pascal}\) & \makecell{$-26.05 \xi^2 + 7.564 \xi - 0.4866$} & \makecell{$13.43 \xi^2 - 0.9107 \xi - 0.0003248$}\\
\hline
\(\tau = \SI{100}{\mega\pascal}\) 1\textsuperscript{st} jump & \makecell{$-240 \xi^2 + 58 \xi - 2.1$} & \makecell{$320 \xi^2 - 1.743 \times 10^{-31} \xi + 1.602\times10^{-17}$}\\
\hline
\(\tau = \SI{100}{\mega\pascal}\) 2\textsuperscript{nd} jump & \makecell{$-320 \xi^2 + 364 \xi - 137.9$} & \makecell{$320 \xi^2 + 324 \xi + 47$}\\
\hline
\(\tau = \SI{120}{\mega\pascal}\) 1\textsuperscript{st} jump & \makecell{$-560 \xi^2 + 54 \xi - 0.9$} & \makecell{$160 \xi^2 - 8.716 \times 10^{-32} \xi + 8.012 \times 10^{-18}$}\\
\hline
\(\tau = \SI{120}{\mega\pascal}\) 2\textsuperscript{nd} jump & \makecell{$-400 \xi^2 + 422 \xi - 153.8$} & \makecell{$240 \xi^2 - 246 \xi + 20.5$}\\
\hline
\end{tabular}
}
\label{Potential Energy}
\end{table}

The curvature of $\Delta H$ at these points yields the angular frequencies listed in Table (\ref{table:omegas}).

\begin{table}[H]
\centering
\caption{Angular frequencies at initial ($\omega _A$) and saddle points ($\omega _B$) at each stress value.}
\small
\resizebox{0.55\textwidth}{!}{
\begin{tabular}{c|c|c}
\hline
\text{Shear Stress (\SI{}{\mega\pascal})} & \textbf{\(\omega_{A}\) (\SI{}{\per\pico\second})} & \textbf{\(\omega_{B}\) (\SI{}{\per\pico\second})}  \\
\hline
30 & \num{0.346} & \num{0.423}  \\
\hline
40 & \num{0.257} & \num{0.358}  \\
\hline
100 1\textsuperscript{st} jump  & \num{1.276} & \num{1.106}  \\
\hline
100 2\textsuperscript{nd} jump& \num{1.278} & \num{1.127}  \\
\hline
120 1\textsuperscript{st} jump & \num{0.903} & \num{1.689}  \\
\hline
120 2\textsuperscript{nd} jump & \num{1.198} & \num{1.252}  \\
\hline
\end{tabular}
}
\label{table:omegas}
\end{table}

\begin{figure}[H]
    \centering
    \includegraphics[width=1\linewidth]{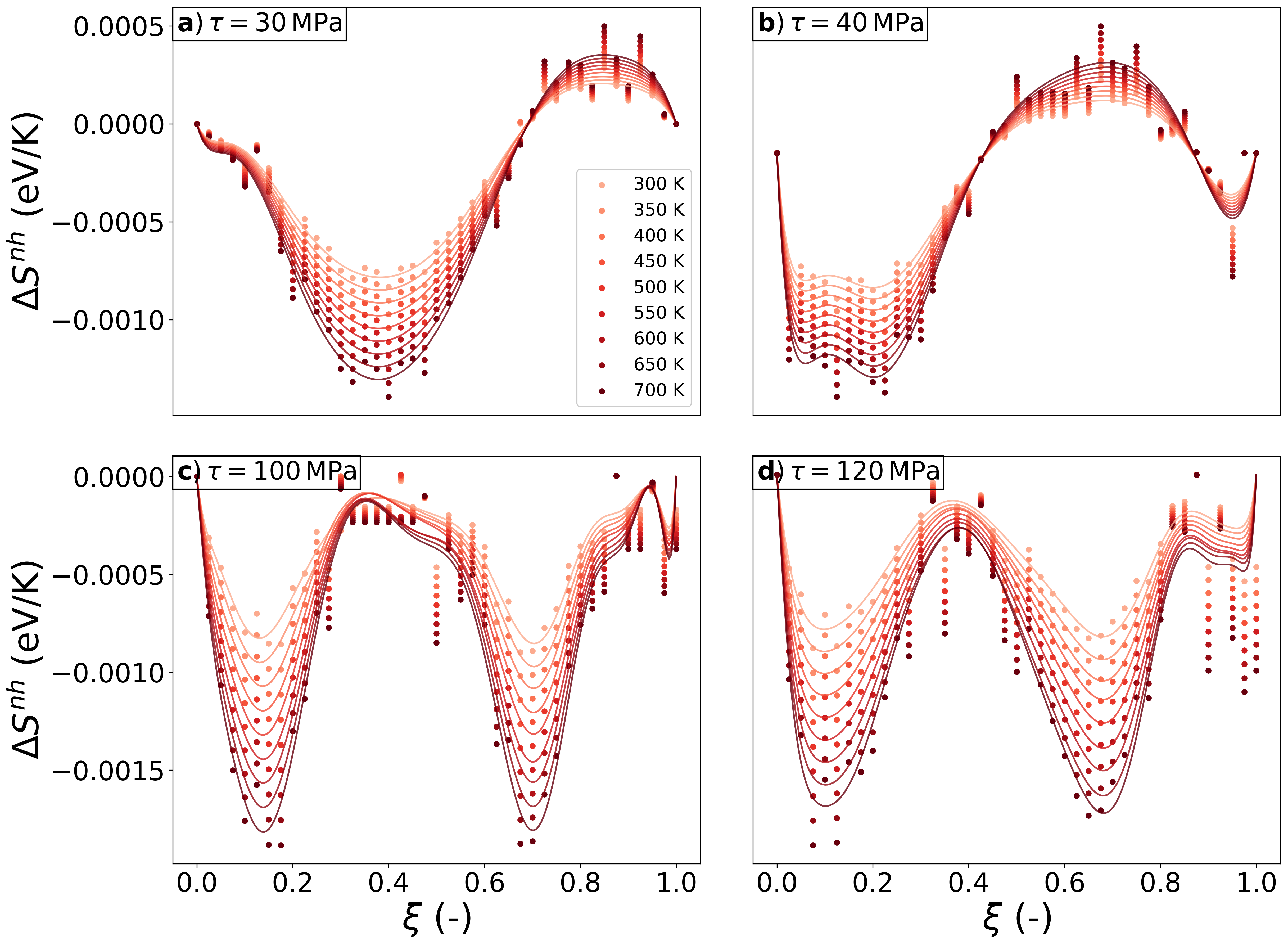}
    \caption{Entropy along the NEB coordinate using Schoeck's method for cell including screw dislocation dipole at a) \SI{30}{\mega\pascal} b) \SI{40}{\mega\pascal}, and edge dislocation dipole at c) \SI{100}{\mega\pascal} and d) \SI{120}{\mega\pascal}}
    \label{fig:entropy}
\end{figure}

Using Eq.\eqref{eq:Schoeck}, we calculated the entropy along the reaction coordinate, as illustrated in Fig. \ref{fig:entropy}, at each replica and with fitted solid lines using splines. For the configuration with a screw dislocation dipole, increasing the temperature leads to a decrease of the minimum and an increase of the maximum entropy values under stress levels of \SI{30}{\mega\pascal} and \SI{40}{\mega\pascal}. A similar trend is observed for the edge dislocation dipole. This temperature-dependent variation in the entropy is primarily attributed to the temperature sensitivity of the elastic constants, as presented in Table \ref{elastic properties}. This temperature dependence suggests that entropy cannot be considered as a constant parameter in Eq.\eqref{eq:TST}. Since there is no hydrostatic force applied to the cell, the first term in Eq.\eqref{eq:Schoeck} should be negligible. Indeed, we computed the results both with and without this hydrostatic term, and it did not affect \(\Delta S\) along the path. We will later discuss in Sec.~\ref{sec:discussion} the implications of having a second-order fit of the elastic constants with temperature and compare with a linear fit resulting in a constant entropy.

Using the entropy and enthalpy values along the MEP, we can compute the rate at which dislocations overcome the barrier imposed by the interaction with the other dislocation of the dipole relying on Kramers rate expression, including the activation free energy (see Fig. \ref{fig:activation energy}). We see that at each temperature, an increase in applied shear stress decreases the activation free energy, which is in agreement with previous studies \cite{Jaime}.  
We show how the free energy profiles change with temperature for different shear stress values shown in Fig. \ref{fig:Free_energy_landscape}:
\begin{figure}[H]
    \centering
    \includegraphics[width=1\linewidth]{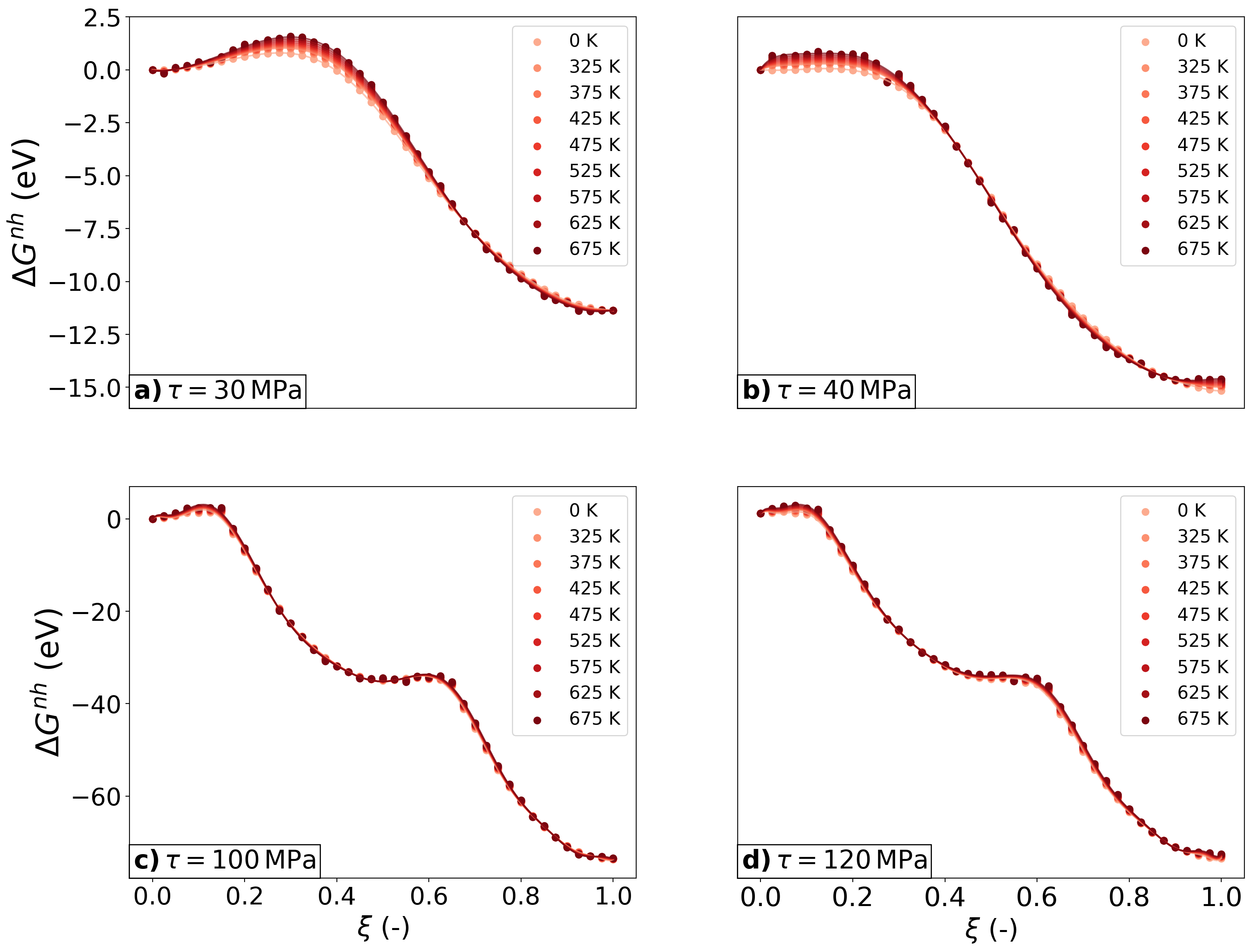}
    \caption{Free energy variation with temperature under different shear stresses a) 30 and b) 40 c) 100, and d)\SI{120}{\mega\pascal} for the configuration including screw and edge dislocation dipoles}
    \label{fig:Free_energy_landscape}
\end{figure}

To validate this approach, we calculated the rates with MD under similar conditions as the inverse of the average waiting times for five independent seeds at various temperatures, stress levels, and friction coefficients. To account for the uncertainty in MD calculations, error bars are added to each MD data point. Figure (\ref{fig:rate vs T inv}) shows that changes in the friction coefficient only shift the rate vertically, indicating that friction does not alter the free energy barrier (slope of the function) but just the prefactor. Theoretically, according to Eq.\eqref{eq:kramers}, friction just affects the attempt frequency. To compare the theory with MD, we use the activation free energies obtained from NEB and Schoeck's approaches, and applied them into Kramers' expression. As shown by the solid lines in Fig. (\ref{fig:rate vs T inv}), the trends of the curves closely match the MD data, with at most an order of magnitude difference at \SI{120}{\mega\pascal} for the edge dislocation, and closer values for the other cases. A key observation in Kramers plots is the change in slope at high temperatures, especially noticeable at higher stress levels for both edge and screw dipole cases. As stress increases and approaches \(\tau_{CRSS}\) at 0 K, the enthalpy barrier decreases, making entropic effects more significant. This leads to a highly temperature-dependent competition between \(\Delta G_a\) and \(k_B T\). The rate decrease with the inverse of temperature follows an Arrhenius relation at low temperatures (as temperature increases the rate increases), although at high temperatures it might transition to a non-Arrhenius behavior (as temperature increases the rate decreases). The model seems to overestimate the MD results, where the non-Arrhenius behavior is not as prominent.  
\begin{figure}[H]
    \centering
    \includegraphics[width=1\linewidth]{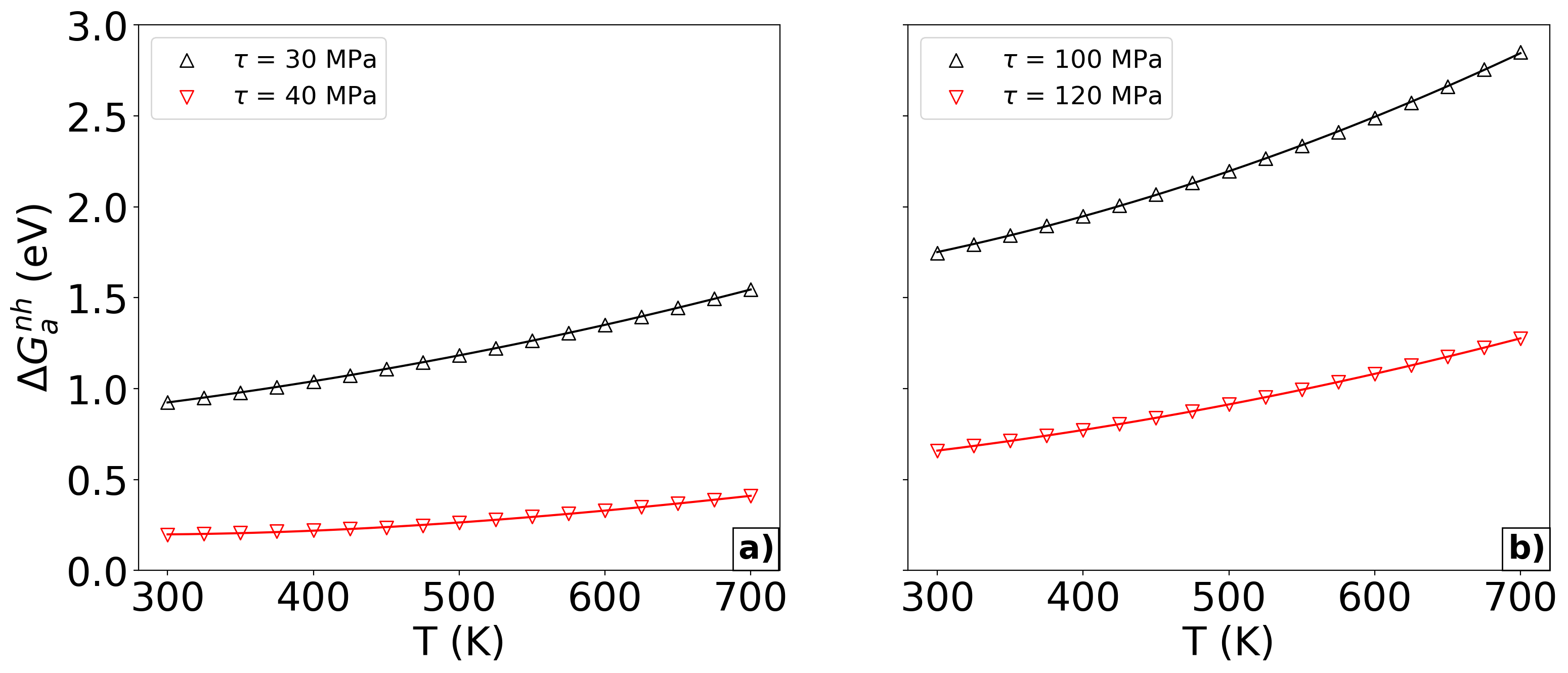}
    \caption{Activation free energy as a function of temperature for cell including a) screw and b) edge dislocation dipole}
    \label{fig:activation energy}
\end{figure}

\begin{table}[H]
\centering
\caption{Activation free energy per as a function of temperature }
\small
\renewcommand{\arraystretch}{1.3} 
\begin{tabular}{c|c}
\hline
\textbf{ Shear Stress (MPa)} & \textbf{\(\Delta G_{eff} (eV)\)}  \\
\hline
\(\tau = \SI{30}{}\)  & $ 1.119 \times 10^{-6} T^2 + 0.00034T + 0.718$ \\
\hline
\(\tau = \SI{40}{}\) & $2.133\times 10^{-6} T^2 - 0.001T + 0.331$ \\
\hline
\(\tau = \SI{100}{}\) & $2.559 \times 10^{-6}T^2 + 1.762 \times 10^{-4} T + 1.468$ \\
\hline
\(\tau = \SI{120}{}\)  & $1.364 \times 10^{-6}T^2 +1.796 \times 10^{-4} T + 0.483$ \\
\hline
\end{tabular}
\label{table2}
\end{table}

\begin{figure}[H]
    \centering
    \includegraphics[width=0.9\linewidth]{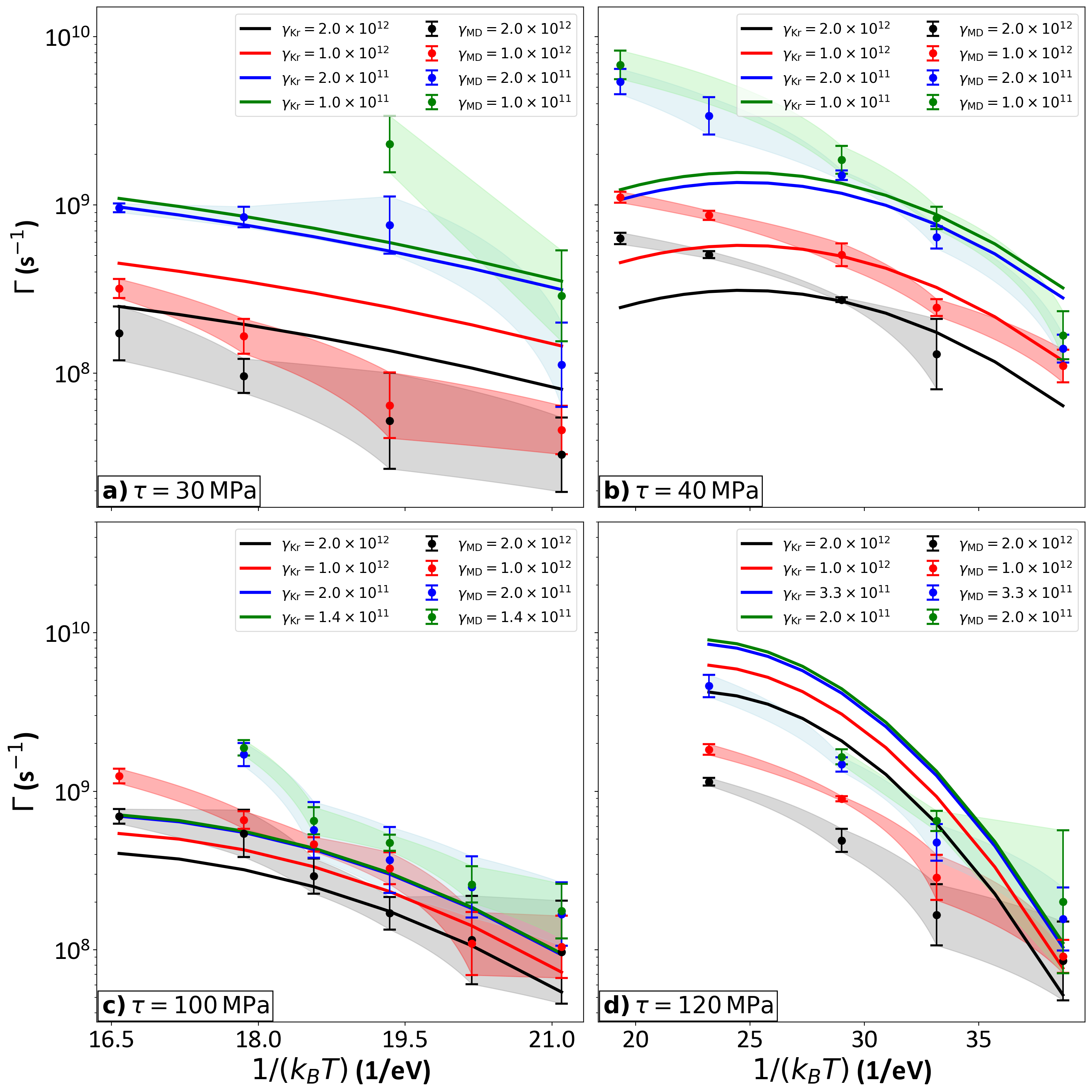}
    \caption{Log-scale rate as a function of temperature inverse for cell including screw dipoles at a) \SI{30}{\mega\pascal} and b) \SI{40}{\mega\pascal} and edge dipoles at c) \SI{100}{\mega\pascal} and d) \SI{120}{\mega\pascal}}
    \label{fig:rate vs T inv}
\end{figure}

\begin{figure}[H]
    \centering
    \includegraphics[width=0.9\linewidth]{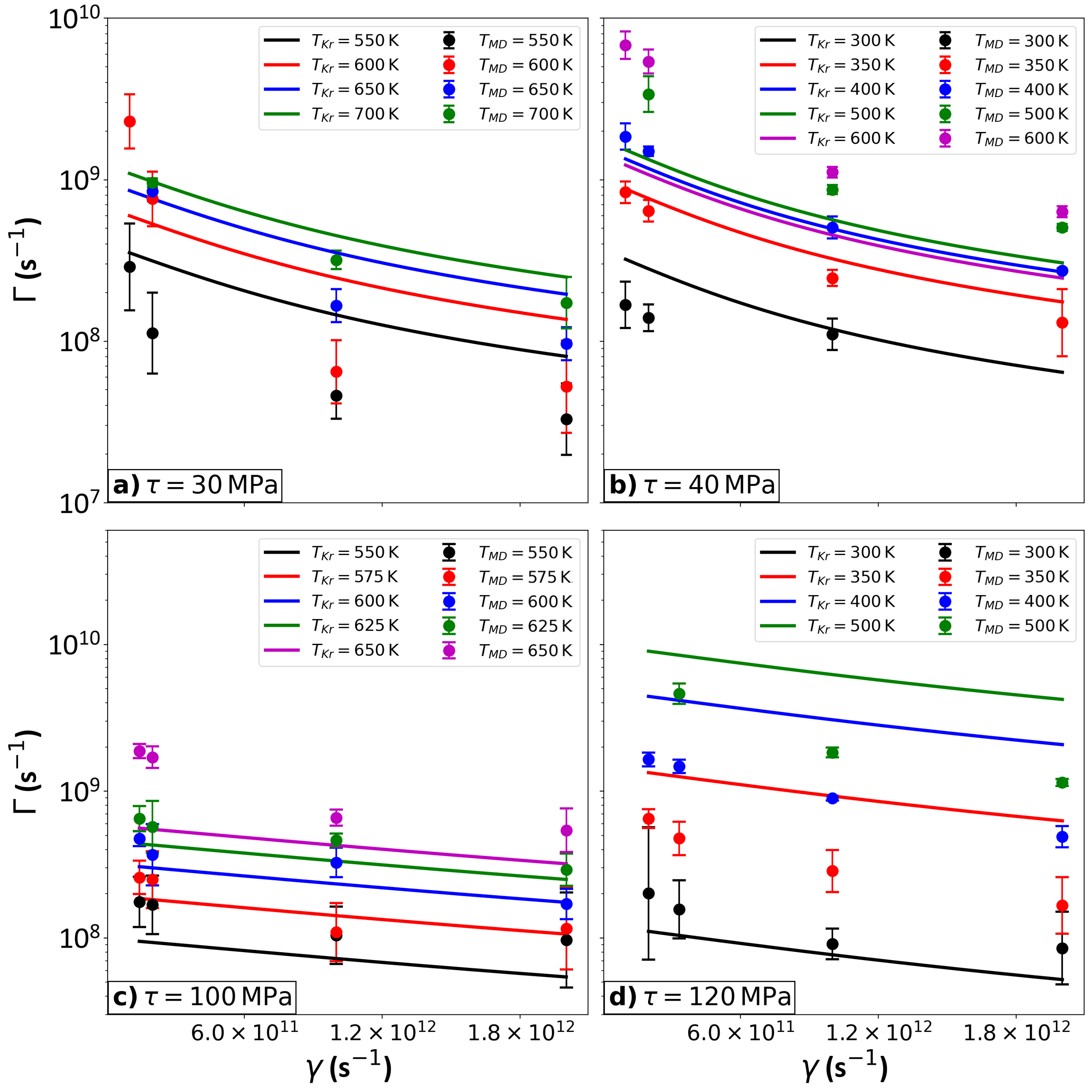}
    \caption{Rate as a function of friction coefficient for a screw dislocation dipole at a) \SI{30}{\mega\pascal} b) \SI{40}{\mega\pascal}, and an edge dislocation dipole at c) \SI{100}{\mega\pascal} and d) \SI{120}{\mega\pascal}}
    \label{fig:rate vs friction}
\end{figure}

To shed more light on the effect of the friction coefficient on the rate values, Fig (\ref{fig:rate vs friction}) compares MD and Kramers rates as a function of $\gamma$. As friction increases, it hampers the movement of the dislocations, resulting in lower rates. In Fig (\ref{fig:rate vs friction}-a) and (\ref{fig:rate vs friction}-b) for a screw dislocation dipole, Kramers leads to a larger curvature of the rates as a function of the friction coefficient compared with Fig (\ref{fig:rate vs friction}-c) and (\ref{fig:rate vs friction}-d) for the edge dipole. The temperature range where we could compute MD rates is \(550 \, \text{K} \leq T_{MD} \leq 700 \, \text{K}\) under \SI{30}{\mega\pascal} and \(300 \, \text{K} \leq T_{MD} \leq 600 \, \text{K}\) under \SI{40}{\mega\pascal} for the screw dipole. Similarly, the results for the edge dipole at \SI{120}{\mega\pascal} are obtained in a lower temperature range (\(300 \, \text{K} \leq T_{MD} \leq 500 \, \text{K}\)) than those for \SI{100}{\mega\pascal} (\(550 \, \text{K} \leq T_{MD} \leq 650 \, \text{K}\)). If the temperature was higher than such limits, the waiting times would not be measurable and the process would not likely be thermally activated any longer, and for lower temperatures, dislocations would not hop over the barrier in a reasonable wall time. Note that the total rates for the edge dislocation cases are computed through the summation of the waiting times for the whole process, involving the jump over two barriers: $t_T=t_1+t_2=\frac{1}{\Gamma_1}+\frac{1}{\Gamma_2}=\frac{1}{\Gamma_T} \rightarrow$ \(\Gamma_T = \frac{\Gamma_1 + \Gamma_2}{\Gamma_1 \times \Gamma_2}\), shown in Fig. (\ref{fig:enthalpy1}).

\clearpage
\section{Discussion}\label{sec:discussion}

\subsection{Implications to the macroscopic response}
In this study, we investigated the effects of temperature, Langevin friction and external shear stress on the rates for dislocation dipoles to overcome the energy barrier imposed by dislocation-dislocation interactions in FCC Cu. Under such conditions, we computed the entropy change induced by internal strains in the system and used different friction coefficients to calculate the rates through Kramers rate theory \cite{kramers1940brownian,hanggi1990reaction}. We obtained the entropy values using Schoeck's formalism along the MEP for the defect to move from initial to final configurations with intermediate states defined utilizing the NEB method with initial linear interpolation. The entropy values for different replicas show an increase in magnitude with temperature due to the temperature dependence of the elastic constants. This leads to a non-negligible entropic effect at high enough temperatures, dominating the rate when the enthalpy barrier reduces as shear stress increases. Fig. (\ref{fig:rate vs T inv}) shows the entropy-dominated region as a deviation from the linear Arrhenius relation mainly at elevated temperatures. This change in behavior is due to the comparable values of the enthalpy barrier (\(\Delta H_a\)), and the entropic barrier (\(T\Delta S_a^{nh}\)). The non-Arrhenius behavior is stronger when the shear stress gets to the vicinity of \(\tau_{CRSS}\) at 0 K, where the enthalpy barrier becomes small. Note that the activation free energy barrier is always positive and significantly larger than $k_BT$.

The activation free energy can be related to the macroscopic deformation through the Orowan model, which relates the dislocation velocity and strain rate:
\begin{equation}
    \dot{\epsilon}=\rho b \bar{v}
    \label{eq: plasticity1}
\end{equation}
where \(\dot{\epsilon}\) is the strain rate, \(\rho\) is the density of mobile dislocations, and \(\bar v \) is the average dislocation velocity. The average velocity can be written as $\bar{v}=H\Gamma_{Kr}$, with $H$ the jump length. Substituting in the Orowan relation we obtain

\begin{equation}
    \dot{\epsilon} =\rho  \nu _{Kr} L H \exp \Big(\frac{-\Delta G_a^{nh}}{k_BT}\Big) =\rho  \nu _{Kr} L H \exp \Big(\frac{-\Delta H_a + T\Delta S_a^{nh}}{k_BT}\Big)
    \label{eq: plasticity2}
\end{equation}

From the data obtained in this work for the configurations including edge and screw dipoles, we find the following polynomial fits to \(\Delta G_a^{nh} (T,\tau)\) 

\begin{table}[H]
\centering
\caption{Activation free energy as a function of temperature and stress}
\small
\renewcommand{\arraystretch}{1.5} 
\begin{tabular}{c|c}
\hline
\textbf{Conf} & \textbf{\(\Delta G_a^{nh}(T,\tau)\)} \\
\hline
Edge & 
\makecell[c]{
  \(1.005 - 7.048 \times 10^{-4}T - 4.836 \times 10^{-6}T^2\) \\
  \(- 8.394 \times 10^{-2} \tau + 6.094 \times 10^{-6}\tau T - 7.423 \times 10^{-8}\tau T^2\)
} \\
\hline
Screw & 
\makecell[c]{
  \(3.314 - 5.287 \times 10^{-4}T + 2.420 \times 10^{-6}T^2\) \\
  \(- 8.346 \times 10^{-2} \tau - 4.274 \times 10^{-6}\tau T - 2.971 \times 10^{-8}\tau T^2\)
} \\
\hline
\end{tabular}
\label{table_crss}
\end{table}

\begin{figure}[H]
    \centering
    \includegraphics[width=0.7\linewidth]{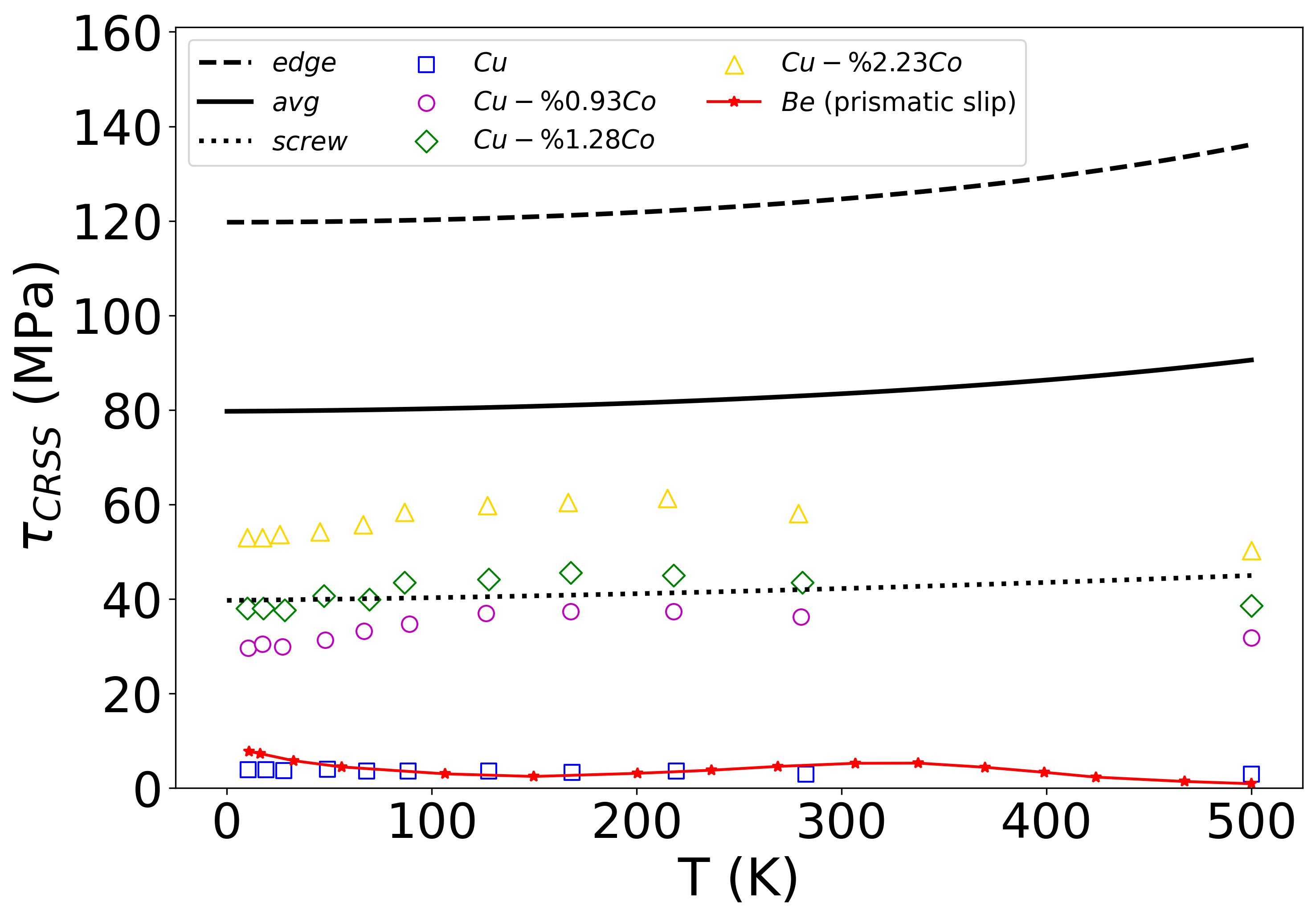}
    \caption{\(\tau_{CRSS}\) resulted from this work (dotted line is for configuration including screw dislocation dipole, dashed line is for the configuration including edge dislocation dipole, and solid line represents the average value of configurations including edge and screw dipoles) compared with prismatic slip in Be, pure Cu, and Cu-Co alloys obtained from experimental observations \cite{SHIM2016276,FUSENIG1993}}
    \label{fig:crss_combined}
\end{figure}

The critical resolved shear stress (CRSS) is defined as the stress that zeros the activation free energy. Hence, we can equate to zero the expressions in Table \ref{table_crss} to obtain the dependence of the CRSS with temperature. The results are shown in Fig.~\ref{fig:crss_combined}, where we observe that there is a slight increase in CRSS with temperature. 
Experimental results \cite{FUSENIG1993} indicate that \(\tau_{CRSS}\) for under-aged Cu and Cu-Au alloys at $T < 200\,\text{K}$ increases as temperature increases, and for $T > 200\,\text{K}$ decreases, with a maximum at around 200 K. It should be noted that this trend change is weaker for pure Cu than its alloys, although the trend seems to persist. Our study sheds light on this anomalous behavior with the role of entropy. At each stress, by changing the temperature, the only property that changes the activation free energy is the activation entropy. The negative activation entropy increases the activation free energy, which leads to an increase in \(\tau_{CRSS}\). Fig. \ref{fig:crss_combined} shows the \(\tau_{CRSS}\) computed in this study compared with available data in the literature for the dependence of the flow stress with temperature. The figure shows that both theoretical and experimental results follow the same increasing trend of stress with temperature.

The differences between approaches can be due to several factors, such as the usual constraints in atomistic modeling, i.e., issues with the interatomic potentials to predict accurate elastic constants, periodic boundary conditions, and the reduced time-scale. One important limitation in our study is that we focus on specific configurations, with just a dislocation dipole in the simulation box. Such configurations might not be representative of the dislocation ensembles present in experiments, but offer a wealth of information for the specific mechanism studied. To explain the flow stress anomaly, Mordike and Haasen \cite{Mordike1962} hypothesized that impurities are most likely responsible for forming obstacles causing the thermally activated anomaly to the flow-stress in the low temperature regime. In Cu-Co alloys, dynamic dislocation effects, hindering kinetic energy-assisted obstacle overcoming were suggested to explain the temperature dependence of the CRSS (Fig.\ref{fig:crss_combined}). This contrasts with higher temperatures, where overdamped motion eliminates such effects \cite{FUSENIG1993}. In hexagonal close-packed (HCP) beryllium, the CRSS for prismatic slip peaks between \SIrange{170}{450}{\kelvin} supposedly due to thermally activated cross-slip forming immobile salient points, which hinder dislocation glide below \SI{300}{\kelvin} but become glissile at higher temperatures \cite{BEUERS1987}.
The yield stress anomaly in Cu and Ni-based intermetallics has also been attributed to diffusion and cross-slip processes  \cite{CAILLARD2001}. In our study of pure Cu, these effects are not present, which might explain why a maximum is observed in experiments while it is not captured in our approach. Despite its limitations, the theoretical results shown in this study hint, for the first time, to an entropic effect as responsible for the anomalous flow stress temperature dependence experimentally observed in Cu and Cu alloys, and in other FCC metals.

\begin{equation}
    \tau_{CRSS_{edge}}(T) = \frac{1.005 - 7.048 \times 10^{-4} T - 4.836 \times 10^{-6} T^2}{8.394 \times 10^{-2} - 6.094 \times 10^{-6} T + 7.423 \times 10^{-8} T^2}
    \label{eq:crss_edge}
\end{equation}

\begin{equation}
    \tau_{CRSS_{screw}}(T) = \frac{3.314 - 5.287 \times 10^{-4} T + 2.420 \times 10^{-6} T^2}{8.346 \times 10^{-2} + 4.274 \times 10^{-6} T + 2.971 \times 10^{-8} T^2}
    \label{eq:crss_screw}
\end{equation}
\vspace{10mm}

\subsection{Implications of rate dependence on friction}
We observe that the dependence of the rates for dislocation-obstacle bypass on the Langevin friction is significant. The glide of dislocations is a dissipative process, which means that the work of the applied stress will convert into heat that will increase the temperature of the system if running in an NVE ensemble. Such increase in temperature will modify the rates for the process involving dislocation motion. On the other hand, thermostats modify the phonon distribution adding scattering that interacts with the moving dislocation. Therefore, the analysis of the dislocation motion under constant temperature conditions requires a deep understanding of the role of the thermostat. 

Under each stress and temperature value studied in this work, we see in Fig.~(\ref{fig:rate vs friction}) that an increase in the friction coefficient hinders dislocation motion, resulting in a lower rate to overcome the barrier. Although the MD simulations capture thermal activation processes in a rather narrow temperature range, the estimates from Kramers approach show good agreement with the MD results. Also, looking at the MD data, we clearly see that friction does not change the energy barrier (slope of the curves), it just modifies the attempt frequency, in full agreement with Kramers expression. The differences in the rates between Kramers and MD might come from uncertainties in the enthalpy barriers, the calculation of atomic strains or the dependence of the elastic constants with temperature. In any case, the trends are fully recovered and the differences in absolute values are significantly lower than an order of magnitude, which indicates that the physical mechanisms are well-captured by the model. 

It is worth noting that, in the regime of frictions explored in this work, we do not observe a plateau in the rates \cite{hanggi1990reaction} as a function of friction, as it is usually the case in solid state processes. Therefore, there is no well-defined rate that can be computed independently from friction. One potential reason for this behavior could be the MD limitation in simulated times, that constrain the different friction values that can be used to estimate rates for these processes. As shown in Fig.~\ref{fig:rate vs friction}, the range of frictions that were probed is rather limited. One potential solution is to extrapolate the rates to zero friction, which would correspond to the transition state theory rate and will be the thermodynamic limit for an infinite system.


\subsection{Meyer-Neldel Rule}

Returning to the third-order Taylor expansion of the specific free energy around temperature (\(\theta = T- T_0\)) and strain (\(V_{ab}\)) within Schoeck's formalism in  Ref.\cite{schoeck1980entropy} we have:
\begin{equation}
\begin{split}
   f = & f_{0} + f_{T} \theta + f_{ik} V_{ik} + \frac{1}{2} f_{T,T} \theta^{2} + f_{T,ik} \theta V_{ik} + \frac{1}{2} f_{ik,lm} V_{ik} V_{lm} + \frac{1}{6} f_{T,T,T} \theta^{3}\\
   & + \frac{1}{2} f_{T,T,ik} \theta^{2} V_{ik} + \frac{1}{2} f_{T,ik,lm} \theta V_{ik} V_{lm} + \frac{1}{6} f_{ik,lm,pq} V_{ik} V_{lm} V_{pq}
\end{split}
\label{eq:helmholtz}
\end{equation}

From Maxwell relations we know that at constant temperature, the specific enthalpy can be written as:

\begin{equation}
    \Delta h = \Delta f + T \Delta s + \Delta e_p + \Delta e_{el} + \Delta e_{SF}
    \label{eqb2}
\end{equation}
where \(\Delta s = -f_{T,ik} V_{ik} - \frac{1}{2} f_{T,ik,lm} V_{ik} V_{lm}\), \(\Delta e_p = \tau_s : ( \vec{b} \otimes \vec{n} ) \rho D\) is the plastic work per unit volume done by the dislocations from the initial configuration to the saddle point, $\rho$ the dislocation density and D its displacement, \(\Delta e_{el} =   \tilde{\tau}_s \tilde{\epsilon}_s - \tilde{\tau}_i \tilde{\epsilon}_i \) is the elastic energy density as a function of stress and strain at saddle and initial states \cite{NAHAVANDIAN2024112954}, and \(\Delta e_{SF}\) is the change in stacking fault energy density between initial and saddle point configurations. In the processes studied here, the last two terms are negligible compared to the plastic energy term. Based on the Meyer-Neldel (MN) compensation rule \cite{meyer1937relation}, we can define \(T_{MN}\) to relate the activation entropy and the activation enthalpy at $T=T_0$:

\begin{equation}
    T_{MN} = \frac{\Delta H}{\Delta S} = \frac{\Delta F}{\Delta S} + T_0 + \frac{\Delta E_p}{\Delta S} =  \frac{T_0 \Delta S + \Delta F +\Delta E_p}{\Delta S}
    \label{eqb3}
\end{equation}

For $\theta \approx 0$ and substituting $\Delta F$ integrating Eq.~\ref{eq:helmholtz} and $\Delta S$ from Eq.~\ref{eq:Schoeck}, we obtain:

\begin{equation}
\begin{split}
  T_{MN} =   \frac{ \int_{\Omega}(-T_0f_{T,ik} V_{ik} - \frac{1}{2} T_0 f_{T,ik,lm} V_{ik} V_{lm} + f_{ik} V_{ik} + \frac{1}{2} f_{ik,lm} V_{ik} V_{lm} + \frac{1}{6} f_{ik,lm,pq} V_{ik} V_{lm} V_{pq} + \Delta e_p)d\Omega}{\int_{\Omega}(-f_{T,ik} V_{ik} - \frac{1}{2} f_{T,ik,lm} V_{ik} V_{lm})d\Omega} 
\end{split}
\label{eqb4}
\end{equation}
Considering the unstressed condition as the reference configuration leads to $f_{ik}=0$. Moreover, following Schoeck, we know \(f_{T,ik} = -f_{ik,lm} \alpha_{lm}\). Further, for cubic materials \(f_{ik,lm} \alpha_{lm} = \alpha_V K \delta _{lm}\) (Eq.(4) in Ref.\cite{schoeck1980entropy}), where $K$ is the bulk modulus and $\alpha _V$ the volumetric thermal expansion coefficient, and \(f_{ik,lm} = C_{iklm}\). Hence
\begin{equation}
\begin{split}
  T_{MN} = \frac{\int_{\Omega}(\alpha_V T_0 K V_{ii} - \frac{1}{2} T_0\frac{\partial C_{iklm}}{\partial T} V_{ik} V_{lm} + \frac{1}{2} C_{iklm} V_{ik} V_{lm} + \frac{1}{6} \frac{\partial C_{iklm}}{\partial V_{pq}}V_{ik} V_{lm} V_{pq} + \Delta e_p) d\Omega}{\int_{\Omega}(\alpha_V K V_{ii} - \frac{1}{2} \frac{\partial C_{iklm}}{\partial T} V_{ik} V_{lm})  d\Omega}
\end{split}
\label{eqb6}
\end{equation}

\begin{equation}
\left\{
\begin{array}{ll}
   \text{a)}\quad  \alpha_V \to 0: & T_{MN} \approx \frac{\int_{\Omega}( - \frac{1}{2} T_0\frac{\partial C_{iklm}}{\partial T} V_{ik} V_{lm} + \frac{1}{2} C_{iklm} V_{ik} V_{lm} + \frac{1}{6} \frac{\partial C_{iklm}}{\partial V_{pq}}V_{ik} V_{lm} V_{pq} + \Delta e_p) d\Omega}{\int_{\Omega}( - \frac{1}{2} \frac{\partial C_{iklm}}{\partial T} V_{ik} V_{lm})  d\Omega} \\\\[10pt]
   \text{b)}\quad \frac{\partial C_{iklm}}{\partial T} \to 0: & T_{MN} \approx \frac{\int_{\Omega}(\alpha_V T_0 K V_{ii}  + \frac{1}{2} C_{iklm} V_{ik} V_{lm} + \frac{1}{6} \frac{\partial C_{iklm}}{\partial V_{pq}}V_{ik} V_{lm} V_{pq} + \Delta e_p) d\Omega}{\int_{\Omega}(\alpha_V K V_{ii} )  d\Omega} \\\\[10pt]
   \text{c)}\quad \frac{\partial C_{iklm}}{\partial V_{pq}} \to 0: & T_{MN} \approx \frac{\int_{\Omega}(\alpha_V T_0 K V_{ii} - \frac{1}{2} T_0\frac{\partial C_{iklm}}{\partial T} V_{ik} V_{lm} + \frac{1}{2} C_{iklm} V_{ik} V_{lm} + \Delta e_p) d\Omega}{\int_{\Omega}(\alpha_V K V_{ii} - \frac{1}{2} \frac{\partial C_{iklm}}{\partial T} V_{ik} V_{lm})  d\Omega}   
\end{array}
\right.
\label{eq:TMN_cases}
\end{equation}

\begin{figure}[H]
    \centering
    \includegraphics[width=0.7\linewidth]{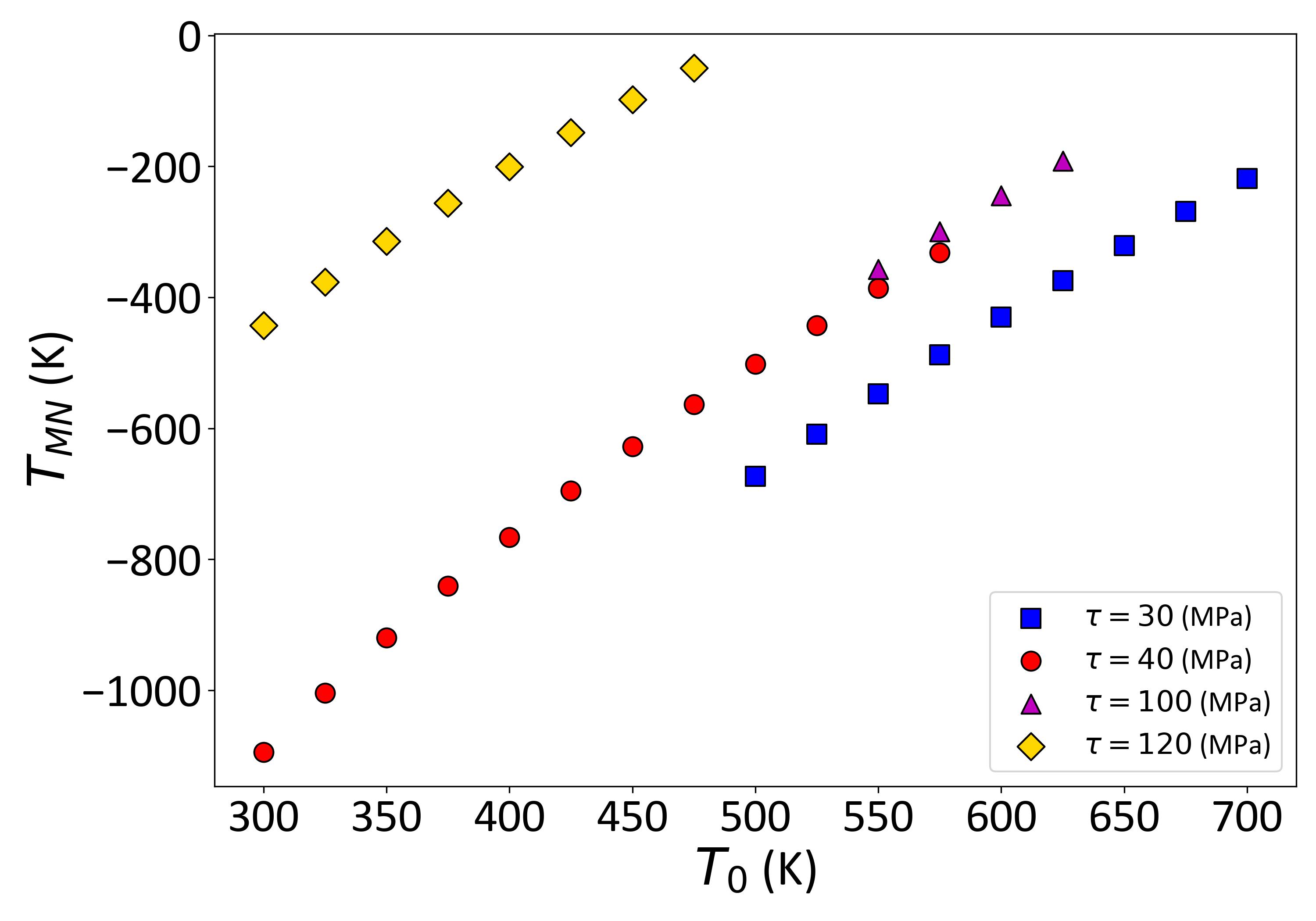}
    \caption{Meyer Neldel temperature at saddle point versus actual temperature for different shear stress values}
    \label{fig:Meyer_Neldel}
\end{figure}

For the cases studied in this paper (Eq.\ref{eq:TMN_cases}-c) \(f_{ik,lm,pq} = 0\) because the elastic constants are only a function of temperature, we can estimate the value of \(T_{MN}\) numerically at different shear stresses. Figure (\ref{fig:Meyer_Neldel}) shows that the activation entropy by MN is negative in all cases in the temperature range used, which leads to the increase in activation free energy with temperature.

\subsection{Rates with constant Entropy}

\begin{figure}[H]
    \centering
    \includegraphics[width=0.9\linewidth]{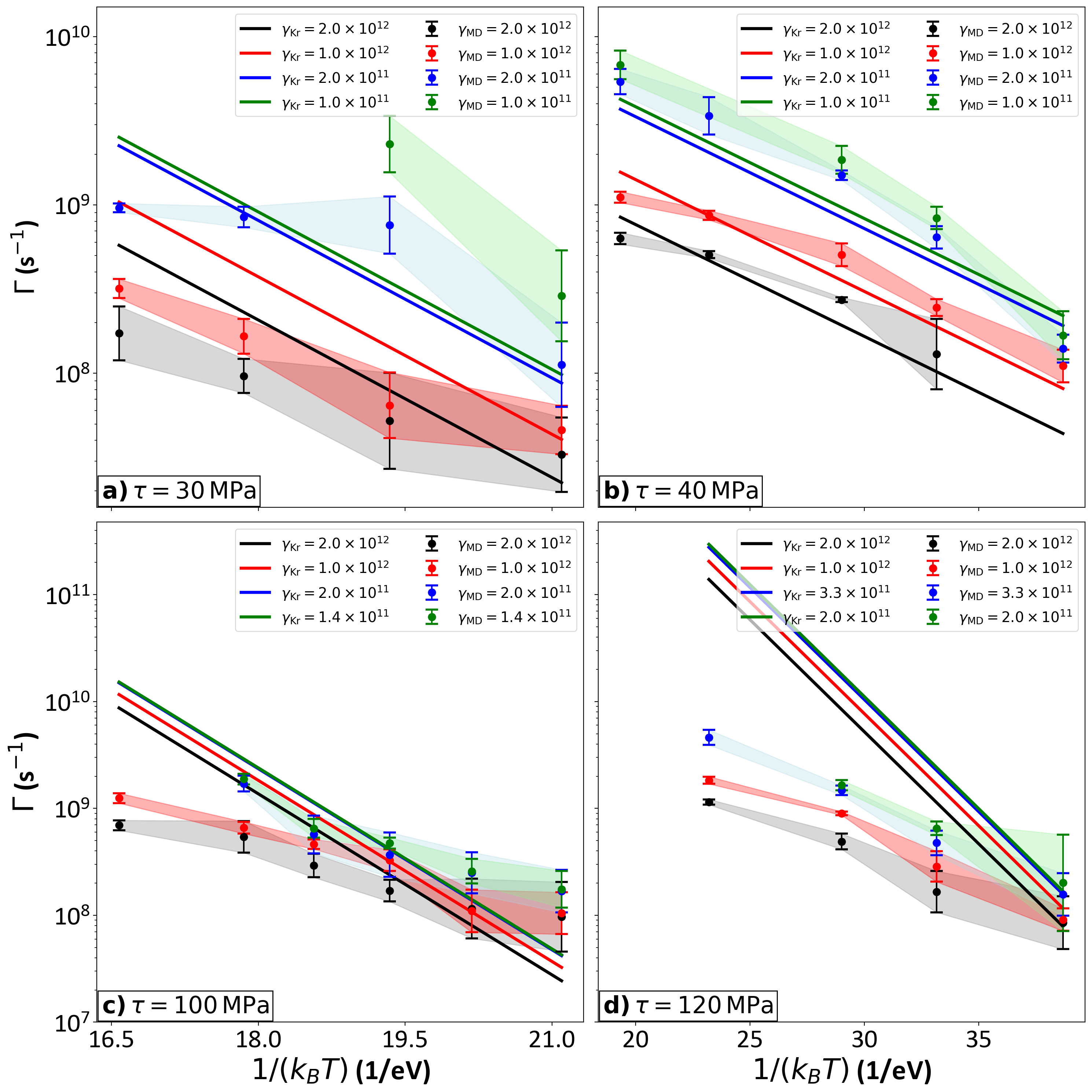}
    \caption{Log-scale rate as a function of inverse temperature for a sample including a screw dipole using a linear temperature dependence of the elastic constants at a) \SI{30}{\mega\pascal} and b) \SI{40}{\mega\pascal} and an edge dipole at c) \SI{100}{\mega\pascal} and d) \SI{120}{\mega\pascal}}
    \label{fig:rate vs T inv linearc}
\end{figure}

To check the effect of the temperature dependence of the entropy, we employ a linear fit to the elastic constants on temperature, which results in constant activation entropy based on Schoeck's formalism (Eq.(\ref{eq:Schoeck})). Comparing Fig.~(\ref{fig:rate vs T inv linearc}) with Fig.~(\ref{fig:rate vs T inv}) we observe that although the Kramers rates are consistent with MD, the non-Arrhenius behavior at high temperature is not captured and it deviates further from the MD values. This deviation is quantified using the mean-squared log error and showed in Table~\ref{tab:rate_compare_MSLE}, which demonstrates that for the majority of the cases, the sum of squared differences between Kramers and MD is larger when rates are computed using constant entropy compared to temperature-dependent entropy.

\begin{table}[H]
\centering
\caption{Mean-squared log error (MSLE) between Kramers and MD rates for each stress and friction coefficient using temperature dependent and constant entropy contributions to the rates. }
\renewcommand{\arraystretch}{1}
\begin{tabularx}{\textwidth}{c|c|c|c}
\hline
\textbf{\(\tau\) (MPa)} & \textbf{\(\gamma\) (s$^{-1}$)} & \textbf{MSLE for \(\Delta S(T)\)} & \textbf{MSLE for \(\Delta S = cte.\)} \\
\hline
\multirow{4}{*}{30} & $2 \times 10^{12}$ & $0.1109$ & $0.1190$\\
                    & $1 \times 10^{12}$ & $0.1791$&$0.1364$ \\
                    & $2 \times 10^{11}$ &$0.0564$ & $0.0754$\\
                    & $1 \times 10^{11}$ &$0.1749$ &$0.4463$ \\
\hline
\multirow{4}{*}{40} & $2 \times 10^{12}$ &$21.0485$ & $19.9884$\\
                    & $1 \times 10^{12}$ &$11.6144$ &$10.9686$ \\
                    & $2 \times 10^{11}$ & $12.7694$& $12.0179$\\
                    & $1 \times 10^{11}$ & $26.739$& $26.8213$\\
\hline
\multirow{4}{*}{100} & $2 \times 10^{12}$ &$0.0293$ & $0.3285$\\
                     & $1 \times 10^{12}$ &$0.0410$ & $0.2583$\\
                     & $2 \times 10^{11}$ & $0.0699$& $0.1138$\\
                     & $1.4 \times 10^{11}$ &$0.0870$ & $0.1093$ \\
\hline
\multirow{4}{*}{120} & $2 \times 10^{12}$ & $0.2732$& $1.6402$\\
                     & $1 \times 10^{12}$ & $0.2081$& $1.5243$ \\
                     & $3 \times 10^{11}$ & $0.1199$& $1.1856$\\
                     & $2 \times 10^{11}$ & $0.1165$& $0.4691$\\
\hline
\end{tabularx}
\label{tab:rate_compare_MSLE}
\end{table}


\section{Conclusion}
In this work, we have analyzed the effect of the Langevin friction and activation entropy on the rate for dislocation dipoles to overcome their dislocation elastic interaction barrier in Cu. We have used two different approaches to compute the rates: 1) direct molecular dynamics (MD) simulations, and 2) Kramer's transition state theory. To estimate the entropy we relied on Schoeck's formalism, while the activation enthalpy was computed using the nudged-elastic band approach. We observe that Langevin friction has a significant effect on the bypassing rates, the larger the friction the lower the rates. The agreement between MD and Kramers' results confirms that the friction coefficient affects only the prefactor and does not modify the energy barrier. MD results seem to show a non-Arrhenius behavior, which is reproduced better when the entropy depends on temperature, a consequence of the higher-than-linear dependence of the elastic constants on temperature. We also showed that the activation entropy plays a significant role in explaining the anomalous increase of flow stress with temperature in metals and alloys.

\clearpage
\section{Declaration of competing interest}
The authors declare that they have no known competing financial interests or personal relationships that could have appeared to influence the work reported in this paper.

\section{Acknowledgment}
The authors acknowledge the support from the US Department of Energy, Office of Science, Office of Fusion Energy Sciences under the grant number \textbf{DE-SC0024515}. 

Additionally, this material is based on work supported by the National Science Foundation under Grant Nos. MRI\# 2024205, MRI\# 1725573, and CRI\# 2010270 for allotment of compute time on the Clemson University Palmetto Cluster.

\newpage
\appendix
\counterwithin{figure}{section} 
\captionsetup[figure]{labelformat=simple,labelfont=bf} 
\renewcommand{\thefigure}{A.\arabic{figure}}

\counterwithin{table}{section}
\captionsetup[table]{labelformat=simple,labelfont=bf}
\renewcommand{\thetable}{A.\arabic{table}}

\section{Elastic Constants and thermal expansion coefficient}
\label{App: elastic_constant}
\begin{figure}[H]
    \centering
    \begin{subfigure}[b]{0.49\textwidth}
        \includegraphics[width=\textwidth]{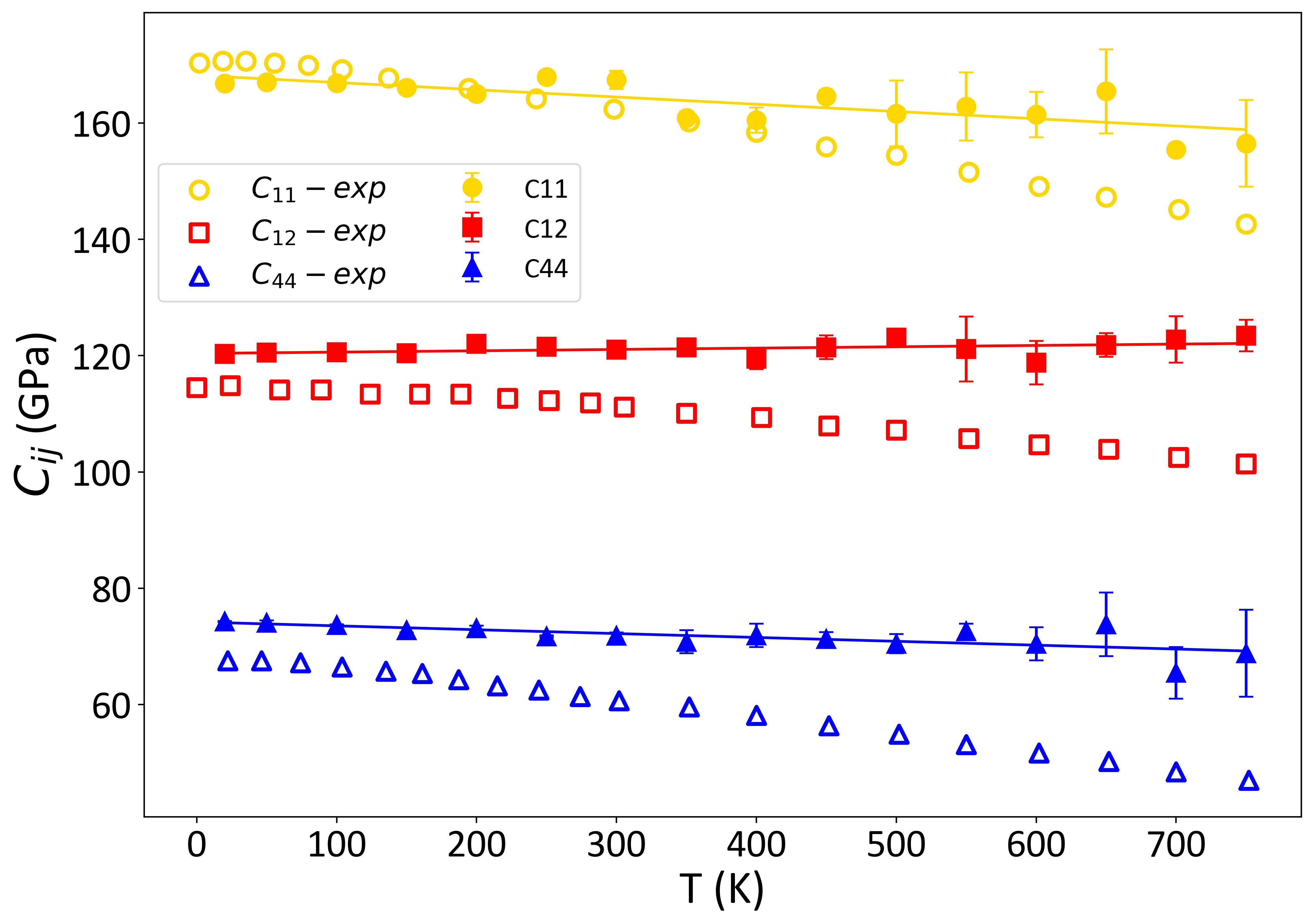}
        \caption{}
        \label{fig:sub1}
    \end{subfigure}
    \hfill
    \begin{subfigure}[b]{0.49\textwidth}
        \includegraphics[width=\textwidth]{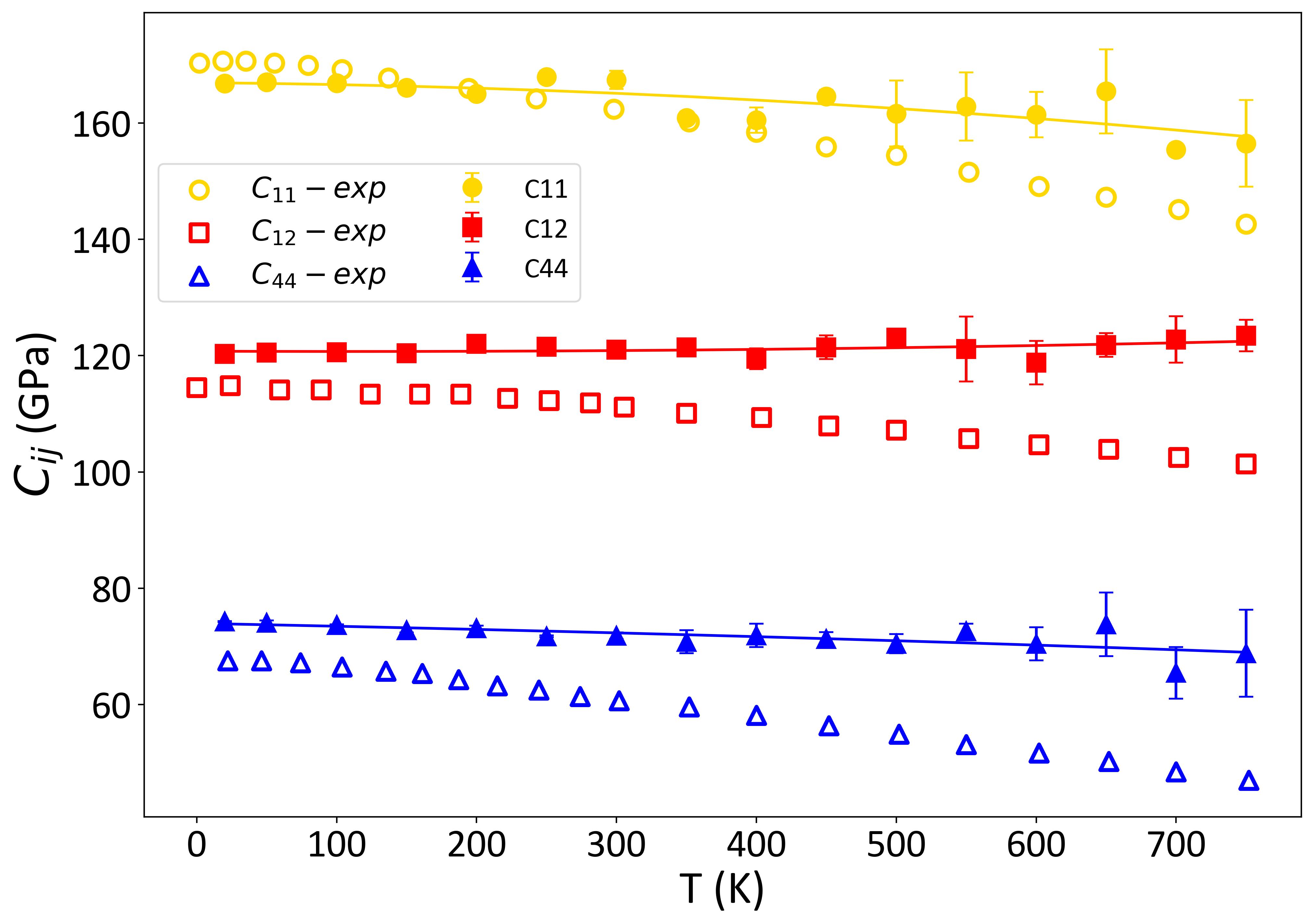}
        \caption{}
        \label{fig:sub2}
    \end{subfigure}
    \caption{\(C_{11}\), \(C_{12}\), and \(C_{44}\) as functions of temperature computed by Y. Mishsin interatomic potential\cite{mishin2001structural}. Experimental data in hollow markers\cite{overton1955temperature} with a) first b) second-order fits}
    \label{fig:elastic constants fig}
\end{figure}

\begin{figure}[H]
    \centering
    \includegraphics[width=0.7\linewidth]{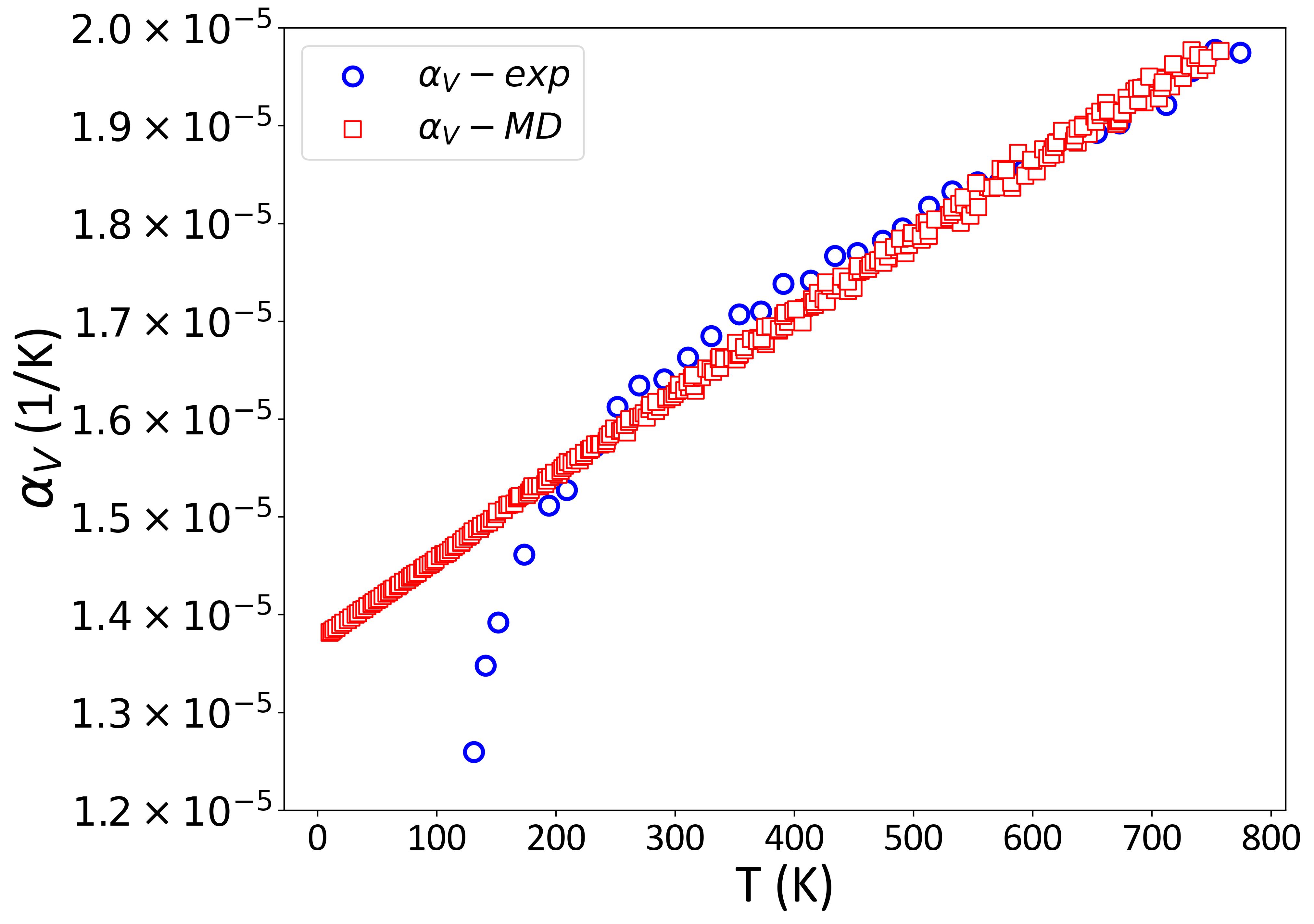}
    \caption{Volumetric thermal expansion coefficient compared with experimental data \cite{exp_Cu_alpha}.}
    \label{fig:alpha_v}
\end{figure}

As it is mentioned in Table \ref{elastic properties}, we have computed the elastic constants relying on Y. Mishin \cite{mishin2001structural} interatomic potentials in the temperature range from 20 to \SI{750}{\kelvin}. Fig (\ref{fig:elastic constants fig}) shows \(C_{11}\), \(C_{12}\), and \(C_{44}\) elastic constants for Cu compared to data in the literature \cite{overton1955temperature}. We observe a decrease of the elastic constants with temperature that aligns with experimental data at low temperatures while it deviates at high temperatures, specifically for \(C_{12}\). The experimental dependence of the elastic constants on temperature is stronger than that obtained experimentally, which will translate in stronger entropic effects than the ones obtained in this work. In addition, computing the \(R^2\) values for first- and second-order fits in temperature shows that the latter explains slightly better the variability of elastic constant data. Also in Fig.~ (\ref{fig:alpha_v}) the thermal expansion coefficient computed by MD is compared with experimental data from the literature \cite{exp_Cu_alpha}. 

\begin{table}[H]
\centering
\caption{\(R^2\) values for \(1^{st}\) and \(2^{nd}\) order fits to elastic constants}
\small
\renewcommand{\arraystretch}{.7} 
\begin{tabular}{c|c|c}
\hline
\textbf{ Elastic constant} & \textbf{\(R^2\) of \(1^{st}\) order fits} & \textbf{\(R^2\) of \(2^{nd}\) order fits}  \\
\hline
\(C_{11}\)  & 0.53 & 0.61 \\
\hline
\(C_{12}\) & 0.22 & 0.27 \\
\hline
\(C_{44}\) & 0.54 & 0.56 \\
\hline
\end{tabular}
\label{R^2 values}
\end{table}

\renewcommand{\thefigure}{B.\arabic{figure}} 
\section{MD waiting time analysis}
\begin{figure}[H]
    \centering
    \includegraphics[width=1\linewidth]{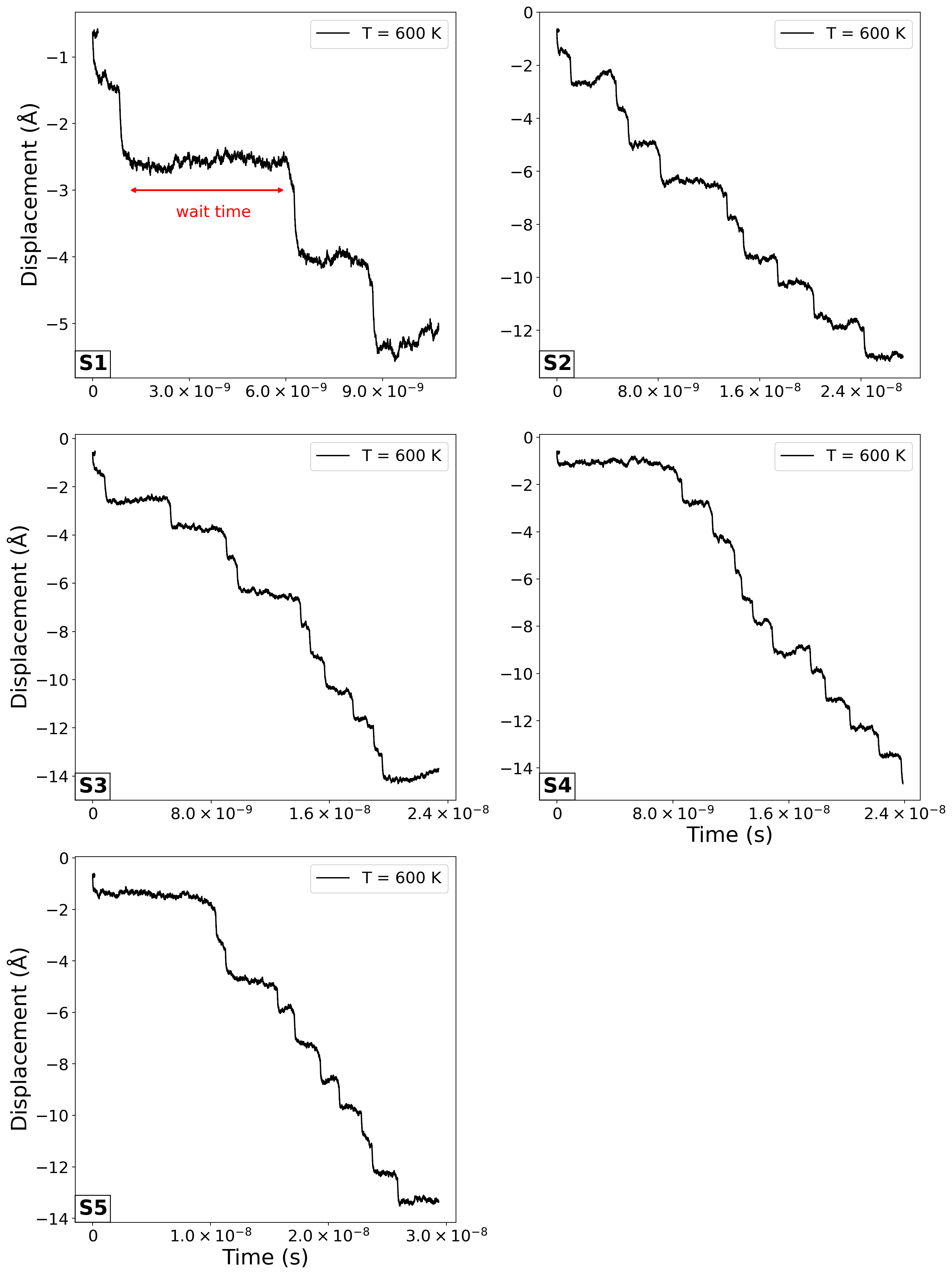}
    \caption{Displacement-time plot for five different realizations under \SI{100}{\mega\pascal}  at \(\gamma = 10^{12} ~(1/s)\) and \SI{600}{\kelvin} for an edge dislocation dipole captured through MD simulations.}
    \label{fig:wait_time_MD}
\end{figure}

Figure (\ref{fig:wait_time_MD}) showcase the displacement-time plots for the case of an edge dislocation dipole with \SI{100}{\mega\pascal} at \(\gamma = 10^{12} ~(1/s)\) and \SI{600}{\kelvin} for five independent realizations. The analysis of the waiting times provides the rates for the dislocations to overcome the free energy barrier induced by the dislocation-dislocation interactions.

 \bibliographystyle{elsarticle-num} 
 \bibliography{refs}

\end{document}